\documentclass[twocolumn,floatfix,superscriptaddress,a4paper,accepted=2025-03-25]{quantumarticle}

\pdfoutput=1
\usepackage{amssymb,amsmath,amstext}
\usepackage{graphicx}
\usepackage{epstopdf}
\usepackage{color}
\usepackage{bm}
\usepackage{appendix}
\usepackage[T1]{fontenc}
\usepackage{bbold}
\usepackage{bbm}
\usepackage{calc}
\usepackage{accents}
\usepackage{latexsym}
\usepackage[colorlinks=true,citecolor=blue,linkcolor=magenta]{hyperref}
\usepackage[numbers,sort&compress]{natbib}

\graphicspath{{Figures/}}

\DeclareRobustCommand{\infig}[1]{%
  \begingroup\normalfont
  \includegraphics[height=\fontcharht\font`\B]{Figures/inline/#1.pdf}%
  \endgroup
}

\newcommand{\dbtilde}[1]{\accentset{\approx}{#1}}

\renewcommand{\vec}[1]{\boldsymbol{#1}}

\begin{document}

\title{Improving the efficiency of learning-based error mitigation}

\author{Piotr Czarnik} 
% \thanks{(ordering/stars tbd).}
\affiliation{Theoretical Division, Los Alamos National Laboratory, Los Alamos, NM, USA.}
\affiliation{Institute of Theoretical Physics, Jagiellonian University, Krakow, Poland.}

\author{Michael McKerns} 
% \thanks{(ordering/stars tbd).}
\affiliation{Information Sciences, Los Alamos National Laboratory, Los Alamos, NM, USA.}

\author{Andrew T. Sornborger} 
% \thanks{(ordering/stars tbd).}
\affiliation{Information Sciences, Los Alamos National Laboratory, Los Alamos, NM, USA.}

\author{Lukasz Cincio}
% \thanks{(ordering/stars tbd).}
\affiliation{Theoretical Division, Los Alamos National Laboratory, Los Alamos, NM, USA.}
\affiliation{Quantum Science Center, Oak Ridge, TN 37931, USA.}

\begin{abstract}
Error mitigation will play an important role in practical applications of near-term noisy quantum computers. Current error mitigation methods typically concentrate on correction quality at the expense of frugality (as measured by the number of additional calls to quantum hardware). To fill the need for highly accurate, yet inexpensive techniques, we introduce an error mitigation scheme that builds on Clifford data regression (CDR). The scheme improves the frugality by carefully choosing the training data and exploiting the symmetries of the problem.  We test our approach by correcting long range correlators of the ground state of XY Hamiltonian on IBM Toronto quantum computer. We find that our method is an order of magnitude cheaper while maintaining the same accuracy as the original CDR approach. The efficiency gain enables us to obtain a factor of $10$ improvement on the unmitigated results with the total budget as small as  $2\cdot10^5$ shots. Furthermore, we demonstrate orders of magnitude improvements in frugality for mitigation of energy of the LiH ground state simulated with IBM's Ourense-derived noise model. 
\end{abstract}
\maketitle

\section{Introduction}
\label{sec:intro}

Quantum computers have already demonstrated quantum advantage over classical computers~\cite{google2019supremacy,zhong2021phase}. Nevertheless, current quantum computers suffer from significant levels of hardware noise which limits their computational power~\cite{wang2020noise,franca2020limitations}. At the same time, high noise levels and limited qubit counts make fault-tolerant quantum computing infeasible~\cite{chen2021exponential}. Consequently, to fully utilize the potential of near-term  quantum computers, methods to approximately correct the effects of noise without performing quantum error correction are necessary~\cite{kandala2018error,larose2022error}. Such methods are called error mitigation methods~\cite{endo2021hybrid}.

Several approaches to error mitigation have been proposed. One of the most popular error mitigation approaches is Zero Noise Extrapolation, which, using a simple regression model, scales the noise strength in a controlled manner and then extrapolates to the zero noise limit~\cite{temme2017error,kandala2018error,dumitrescu2018cloud,otten2019recovering,giurgica2020digital,he2020zero,cai2020multi,kim2021scalable}. Another approach is probabilistic error cancellation which introduces additional gates to the executed circuit~\cite{temme2017error,endo2018practical}. The additional gates are introduced probabilistically according to the device noise model in order to cancel noise effects. An alternative approach is to use multiple copies of a noisy state to prepare its purification~\cite{koczor2020exponential,huggins2020virtual,czarnik2021qubit,koczor2021dominant,huo2021dual, cai2021resourceefficient,seif2022shadow,hu2022logical}. Such an approach is called Virtual Distillation. Other proposals are based on quantum phase estimation~\cite{o2021error}, symmetries of the mitigated system~\cite{mcardle2019error,bonet2018low,otten2019noise,cai2021quantum} or are unified methods bringing together different approaches to  error mitigation~\cite{xiong2021quantum,yoshioka2021generalized, mari2021extending}. Finally, circuit compilation methods can be used to reduce noise severity in combination with the error mitigation methods mentioned above~\cite{cincio2018learning,cincio2020machine,murali2019noise,khatri2019quantum,sharma2019noise}.  Although many of these approaches have had success in small circuits, it remains to be seen which of them prove to be the most powerful in the quantum advantage regime.

Another recently proposed approach is learning-based error mitigation~\cite{czarnik2020error,strikis2020learning}, whose advantages we demonstrate in this work.  Learning-based error mitigation learns the effects of noise from classically simulable quantum circuits which are close to non-simulable circuits of interest. A large class of such approaches evaluates an observable's expectation value for targeted training circuits on a quantum device using training data obtained classically. The training data are fitted with an ansatz capturing a relation between noisy and noiseless  expectation values~\cite{czarnik2020error,montanaro2021error,vovrosh2021simple,urbanek2021mitigating,zhukov2022quantum,liao2023machine}. The ansatz is used to correct the noisy expectation value for the circuit of interest. Learning-based error mitigation methods have been shown to perform well for quantum  circuits  with large qubit counts  and depths~\cite{czarnik2020error,sopena2021simulating}. They have also been shown to outperform other state-of-the-art approaches~\cite{bultrini2021unifying}. The learning-based approach is a flexible framework which enables one to improve mitigation quality by expanding the training data, for example accounting for the effects of varying noise strength on the expectation values of observables~\cite{lowe2020unified,bultrini2021unifying}. These features make it a promising candidate for error mitigation for near-term quantum advantage applications.

One of the main limitations of error mitigation is the required shot cost~\cite{endo2018practical}. In the case of learning-based error mitigation it has been shown that to maximize mitigation quality large shot resources are required~\cite{lowe2020unified, bultrini2021unifying}. The required shot numbers seriously restrict the practical potential of most powerful and sophisticated approaches. Furthermore, instability in a quantum computer configuration that takes place over time introduces further limitations on the applicability and corrective power of error mitigation.  Since quantum computers are effectively experimental apparatuses whose control parameters drift over time,  training circuit data acquired over long times using many shots may not accurately reflect noise occurring during the execution of a circuit of interest. We show here that this effect is indeed strong enough to seriously affect error mitigation performance. Finally, some of the most promising near-term applications like variational quantum algorithms require measurement of many quantum circuits to find a final circuit that forms a solution to the problem~\cite{endo2021hybrid, cerezo2020variationalreview}. To apply error mitigation  methods to such problems one needs to optimize the shot-efficiency of the error mitigation approach~\cite{wang2021can,guo2024experimental}.

In this work, we identify a major problem limiting the shot-efficiency of learning-based error mitigation. One  of the most common ways to generate the training data is to use near-Clifford circuits that are similar to the circuit of interest. Such circuits are obtained by substituting non-Clifford gates in the circuit of interest by Clifford gates. We show that such a procedure leads to training circuits for which both noisy and exact expectation values of the observable of interest cluster around $0$. We show that this effect leads to a large increase in the shot number necessary to learn the effects of noise on the observable of interest.

We  propose two strategies to prevent clustering and improve the shot-efficiency of learning based error mitigation. First, we propose to generate near-Clifford training circuits with widely spread distributions of noisy and exact expectation values of the observable of interest. We propose algorithms to generate such distributions which do not require quantum resources (shots) and therefore are suitable for improving the shot-efficiency.  

\begin{figure}[ht!]
\includegraphics[width=0.95\columnwidth]{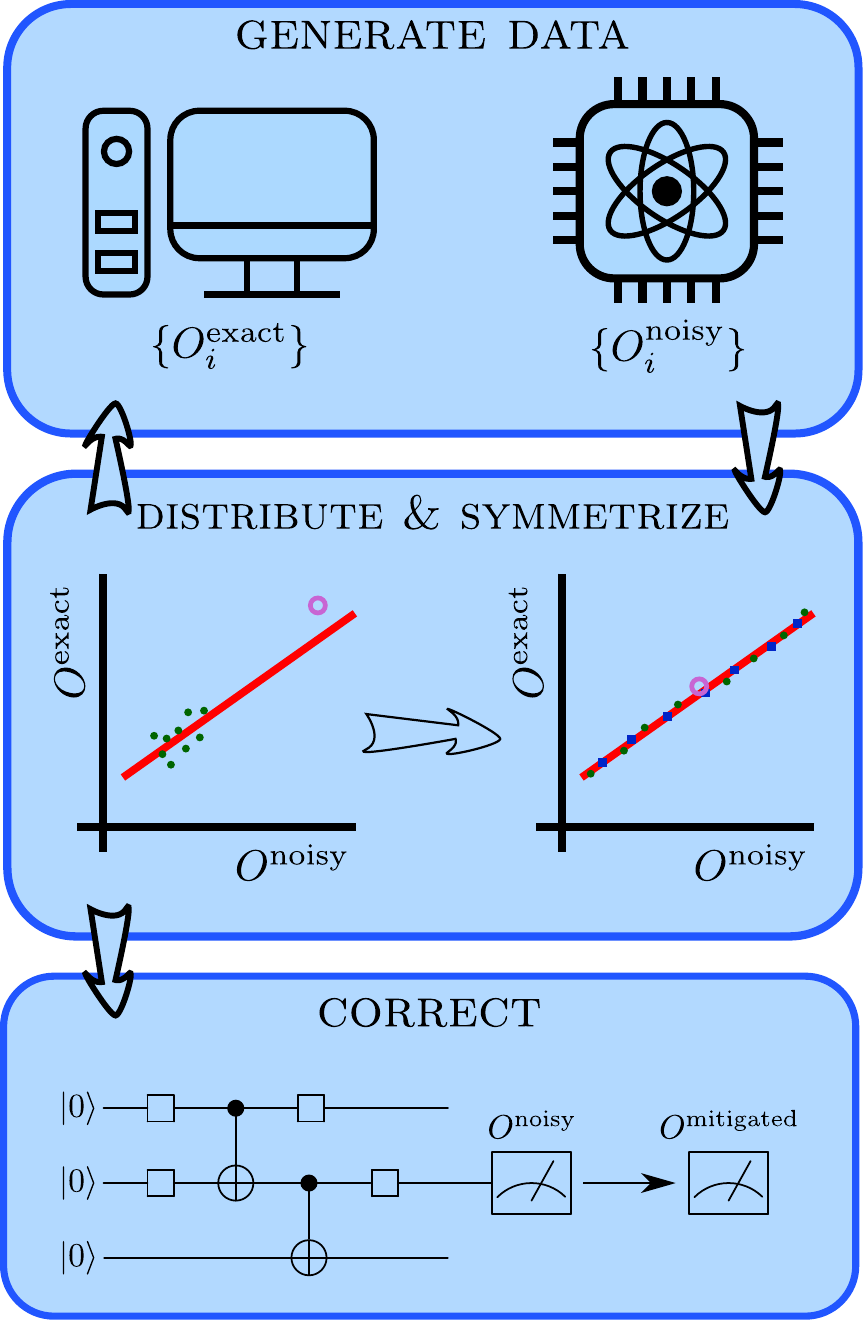}
\caption{ {\bf Improved Clifford Data Regression method.} Circuit of interest  together with observable $O$ are used to generate training data $\{  O^{\rm exact}_i, O^{\rm noisy}_i \}_i$ made out of observables computed from classically simulable circuits similar to the circuit of interest. They are evaluated on both, classical ($O^{\rm exact}_i$) and noisy quantum ($O^{\rm noisy}_i$) computers, as shown in the upper panel.    Previous methods for generating training circuits resulted in data clustering, a situation in which the majority of training circuits share similar $O^{\rm exact}$, as depicted in a cartoon in the left part of the middle panel where  the green circles are the training data and \infig{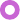} represents $(O^{\rm exact},O^{\rm noisy})$ for the circuit of interest.  This effect necessitates the use of large training data sets to maintain high-quality correction $O^{\rm mitigated}$. We solve that problem by distributing and symmetrizing the training data, as depicted in the right part of the middle panel. That is achieved by generating the training circuits with a large variance of $O^{\rm exact}$ using classical computation.   Our approach significantly reduces the shot cost of data-driven error mitigation techniques while preserving their high correction quality. Symmetries of the problem, if present, can be used to enhance the data further, i.e., to generate additional data (\infig{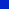}) from initially generated (\infig{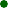}) one, at no extra cost. Finally, a learned ansatz (the red line) is used to correct $O$ for the circuit of interest. 
}
\label{fig:method}
\end{figure}

Second, for circuits with symmetries we propose to use multiple observables, whose expectation values are equal by virtue of the symmetries, to increase the amount of training data available to learn the  ansatz for noiseless expectation values. Such an increase enables us to learn from a fraction of the expectation values which are not clustered around $0$. We show how to use this strategy to improve the shot-efficiency of error mitigation. We summarize schematically the proposed approaches in Fig.~\ref{fig:method}.  To demonstrate the power of these strategies we mitigate correlation functions of the ground state of the XY model prepared using the IBM Toronto quantum computer. We obtain an  order of magnitude improvement in mitigation quality over standard learning-based error mitigation for moderate mitigation shot costs of order $10^5$. Furthermore, for such shot costs, we find that the new method outperforms the standard one even when using a factor of 10 fewer shots.
As a second example of the new approach's usefulness, we mitigate the ground state energy of LiH molecule that does not have exploitable symmetries. In this case, we obtain orders of magnitude shot cost reduction for a real-device-inspired noise model using the first strategy.

In section~\ref{sec:CDR} we describe a learning-based error mitigation method whose shot-effectiveness we improve here. Section~\ref{sec:eff} introduces the clustering problem and the strategies to improve shot-efficiency by alleviating it. Section~\ref{sec:ben} discusses some benchmark results obtained for the IBM Toronto quantum computer and a noise model derived from IBM Ourense by gate tomography. We conclude and discuss possible extensions of this work in Section~\ref{sec:conclusions}. Technical  details of a standard method of near-Clifford training  circuit construction on which we improve here are described in Appendix~\ref{app:tcirc_rand}.  In Appendix~\ref{app:tcirc_sol} we prove the existence of the shot-efficient training circuits for popular variational ansaetze and Pauli observables. There, we also give an efficient semi-analytic procedure to construct a subclass of such circuits.  Details of numerical approaches to construct the shot-efficient training data from more general training circuits are given in Appendix~\ref{app:tcirc_eff}.
Appendix~\ref{app:symmetry} describes the implementation of symmetry constraints on a candidate solution to a minimization problem.
Appendix~\ref{app:model} discusses the IBM Ourense derived noise model. In Appendix~\ref{app:ts_toronto} we expand on the choice of hyperparameters for our IBM Toronto error mitigation implementation. In Appendix~\ref{app:drift} we show evidence of the significant impact of hardware changes over time on performance of error mitigation.  Finally, in Appendix~\ref{app:LiH_Nscal}, we discuss in more detail the effects of the choice of CDR hyperparameters on the mitigation quality.

\section{Clifford Data Regression}
\label{sec:CDR}
Here, we work in the framework of Clifford Data Regression (CDR) \cite{czarnik2020error}. CDR uses classically simulable near-Clifford circuits similar to a circuit of interest to learn a linear ansatz, that is used to mitigate an expectation value of a targeted observable for a circuit of interest. 
We summarize the method by $4$ steps outlined below.

\begin{enumerate}
\item
Create $N_t$ training circuits with $N$ non-Clifford gates substituting non-Clifford gates in the circuit of interest by nearby Clifford gates.  In this work, we additionally allow the non-Clifford  gates to be substituted by similar non-Clifford gates to obtain the training circuits. 
We describe a standard algorithm to generate the training circuits in Appendix~\ref{app:tcirc_rand} and introduce the shot-efficient training circuit generation methods in Section~\ref{sec:eff} providing further details in Appendix~\ref{app:tcirc_eff}.
\item
Evaluate an expectation value of the observable of interest $O$ for each of the training circuits both classically and with a quantum computer. With this procedure, one obtains pairs of noisy $O^{\rm noisy}_i$ and exact expectation values $O^{\rm exact}_i$ for each training circuit.   
\item
Assume a linear relation between $O^{\rm exact}$ and $O^{\rm noisy}$,
\begin{equation}
O^{\rm exact} = a O^{\rm noisy}+b \ ,
\label{eq:CDR}
\end{equation}
where $a$ and $b$ are given by 
\begin{equation}
(a, b) = {\rm argmin} \sum_{i=1}^{N_t} (O^{\rm exact}_i-aO^{\rm noisy}_i-b)^2.
\end{equation}
\item 
Use $a$ and $b$ found in step 3 to mitigate the expectation value $O$, for the circuit of interest 
\begin{equation}
O^{\rm mitigated}_{\rm coi} = a O^{\rm noisy}_{\rm coi} + b.
\end{equation}

\end{enumerate}

CDR assumes that the correction for the circuit of interest can be learned from classically simulable circuits obtained by substituting some of the gates in the circuit of interest with similar gates. 
Furthermore, it assumes that the training data are obtained by measuring an expectation value of the observable of interest. To motivate these assumptions, we consider the case of IBM's quantum computers which are one of the leading approaches to digital quantum computing. They have a native gate set built with $R_Z(\theta)=e^{-i \theta/2 Z}$, $X$, $\sqrt{X}$ and CNOT gates, with $\theta \in [0, 2\pi)$, and $X$, $Z$ being single-qubit Paulis. $X$, $\sqrt{X}$ and CNOT are Clifford gates, while in general $R_Z(\theta)$ is not a Clifford gate.

 Circuits built from Clifford gates and $N$ non-Clifford $R_Z(\theta)$ gates, with $N$ of order of $10$ or smaller, can be currently easily simulated classically~\cite{fast2022pashayan}. Substituting non-Clifford $R_Z(\theta)$  in the circuit of interest with Clifford $R_Z(\theta)$ ($\theta \in \{0, \pi/2, \pi, 3\pi/2\}$), to leave only $N$ non-Clifford $R_Z(\theta)$ gates, we obtain a circuit for which an expectation value of the observable of interest can be computed classically. 

At the same time, $R_Z(\theta)$ gates can be performed `virtually' in a zero-duration mode, consequently being effectively noiseless~\cite{mckay2017efficient}. More importantly, that implies that noise associated with the gate does not depend on the choice of $\theta$, suggesting that the near-Clifford circuits obtained as sketched above can be used as the training circuits. This conjecture is validated by multiple observations that for IBM's quantum computers, effects of the noise on the observables depend strongly on the locality of the observable of interest and number and arrangement of CNOT gates in the circuit of interest~\cite{urbanek2021mitigating,kim2021scalable,tran2023locality}, which in the case of CDR are the same for the training data and the circuit of interest. 

To learn the correction, an ansatz that connects the noisy and the exact expectation values is needed. For the training circuits that are very similar to the circuit of interest and the training data generated with the observable of interest, we expect that a simple ansatz is sufficient. Indeed, CDR successfully uses a linear ansatz~(\ref{eq:CDR}). This choice is motivated by an observation that for the IBM's quantum computers, effects of the noise on the observable of interest can be approximated by an effective global depolarizing noise model~\cite{czarnik2020error,vovrosh2021simple}, which can be perfectly corrected by the linear ansatz~(\ref{eq:CDR})~\cite{czarnik2020error}.

CDR has been demonstrated to provide order-of-magnitude improvements in expectation values of observables for IBM-hardware experiments and diverse applications~\cite{czarnik2020error,sopena2021simulating,zhang2021variational}, making it a simple yet powerful approach to IBM-hardware noise mitigation.  Below, we discuss the training circuit construction assuming IBM's native gate set for the sake of presentation clarity.  

To apply CDR, one needs to choose a value of the $N$ hyperparameter. As the classical cost of the expectation values simulation grows exponentially with $N$, $N=0$ minimizes the classical overhead. Still, simulations of the near-Clifford circuits with $N=30$ $R_Z(\theta)$ gates can be easily performed with a laptop computer, making minimizing $N$ unnecessary. There are two motivations for using $N>0$. First, it has been originally proposed~\cite{czarnik2020error} to use non-zero $N$ to further bias the training circuits towards the circuit of interest by retaining some of the non-Clifford gates present in the circuit of interest. Along those lines, Ref.~\cite{czarnik2020error} reported modest improvements in the mitigation quality with increasing $N$ for a Variational Quantum Eigensolver application and an IBM-quantum-computer-derived noise model.

Another motivation is that  $N>0$ allows a more diverse set of training circuits. Similarly, allowing for non-Clifford $R_Z$ rotations with angles different than in the circuit of interest further increases a pool of potential training circuits. Therefore, we may be able to better characterize variance of deviations from the CDR ansatz~(\ref{eq:CDR}) among circuits similar to the circuit of interest and leverage that to obtain more robust CDR correction. Here, we use $N=10$ and $N=30$ to balance classical computational costs with the benefits of a more diverse set of training circuits. Analogically, though $N_t=2$ is sufficient to learn a linear ansatz, larger $N_t$ may allow better characterization of the deviations from linear behavior among the near-Clifford circuits. In this work, we explore $N_t \in \{2, \ldots, 100\}$, and typically notice systematic improvements of the mitigation quality with increasing $N_t$ as discussed in Section~\ref{sec:ben}.  

To place CDR within a context of other learning-based error mitigation methods, we note that several application-specific approaches have been proposed that use classically simulable training circuits specific for a given application~\cite{montanaro2021error,zhukov2022quantum}. Furthermore, usage of more restrictive training data obtained only from $N=0$ Clifford circuits has been explored~\cite{urbanek2021mitigating}. Additionally, more general sets of noisy training data with expectation values from multiple noise levels have been applied~\cite{lowe2020unified,bultrini2021unifying}, though those methods have been found to be less shot-efficient than CDR~\cite{lowe2020unified,bultrini2021unifying}. Finally, more general asaetze like neural networks and random forests have been used to correct multiple observables and circuits of interests using a single instance of a trained ansatz~\cite{zhukov2022quantum, liao2023machine}. We note that such approaches are alternative paths to improve the shot-efficiency of learning-based error mitigation~\cite{liao2023machine}.

\begin{figure*}[t]
\begin{center}
\includegraphics[width=0.9\textwidth]{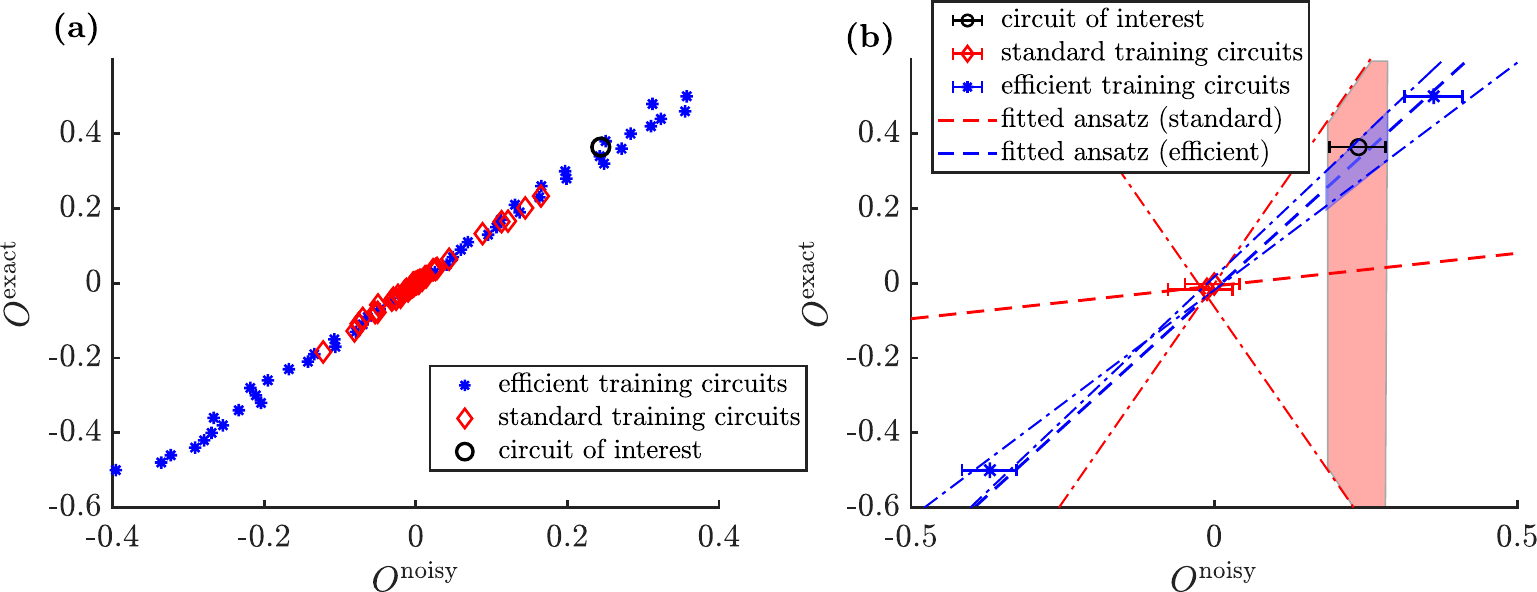}
\end{center}
\caption{ \textbf{Improving  CDR shot efficiency by distributing training circuits.} In \textbf{(a)} we show $O^{\rm exact}$ versus noisy $O^{\rm noisy}$ expectation values of a given observable computed from a circuit of interest as well as from CDR training circuits. The noisy expectation values were obtained in the limit of infinite shot number. We show results for two methods of training circuit construction. The results obtained with the standard method (random substitutions of non-Clifford gates by Clifford ones) are shown with red diamonds. They cluster around a $O^{\rm noisy} = O^{\rm exact} = 0$. The blue asterisks are results obtained for a set of training circuits with a wide distribution of $O^{\rm exact}$ generated by MCMC. This set also prevents clustering of $O^{\rm noisy}$. In \textbf{(b)} we show the results of CDR error mitigation for a small shot cost. We use two training circuits generated by both methods. We evaluate $1000$ shots to estimate $O^{\rm noisy}$ for both training circuits and the circuit of interest. Error bars correspond to 5\textsuperscript{th} and 95\textsuperscript{th} quantiles of $O^{\rm noisy}$ estimated from 100 shot noise instances, while the markers correspond to the median value. We performed CDR fits for all shot noise instances. We show CDR fits corresponding to 50\textsuperscript{th} (dashed lines), 5\textsuperscript{th} and 95\textsuperscript{th} (dashed-dotted lines) quantiles of the mitigated expectation value. The shaded areas show confidence intervals of the mitigated expectation value based on statistical distribution of the  fits. We see that CDR with distributed training circuits results in both smaller uncertainty and better quality of the mitigated expectation values. Thus, these results demonstrate that preventing clustering of the training data indeed improves the shot-efficiency of CDR error mitigation. %  
Here, in both \textbf{(a)} and \textbf{(b)}, the circuit of interest approximates the ground state of an 8-qubit XY model (\ref{eq:H}) and $O=X_1X_5$, where $X$ is a Pauli matrix. To generate $O^{\rm noisy}$, we use in both panels a noise model described in Appendix~\ref{app:model}. 
}
\label{fig:cluster_1obs}
\end{figure*}

\section{Improving efficiency of learning-based error mitigation }
\label{sec:eff}

\subsection{Clustering of near-Clifford training  circuits}
\label{sec:cluster}
The most popular method for choosing training circuits for CDR is to randomly replace most of the non-Clifford gates in the circuit of interest by nearby Clifford gates~\cite{czarnik2020error,larose2022mitiq,cirstoiu2022volumetric}, as discussed in detail in Appendix~\ref{app:tcirc_rand}.  We observe that for circuits of interest with a large number of non-Clifford gates this method results in the clustering of training data caused by the clustering of $O^{\rm exact}$ around $0$.  We visualize  this effect in  Fig.~\ref{fig:cluster_1obs}(a) for a typical training set obtained with a random substitution  method for a half-chain correlator of the ground state of an  8-qubit XY model using a noise model derived from gate tomography of IBM Ourense quantum computer.

As a thought experiment we  consider the case of both training circuits and circuit of interest satisfying the CDR ansatz (\ref{eq:CDR}) perfectly. Then, the clustering of $O^{\rm exact}$  around $0$ implies also  clustering of  $O^{\rm noisy}$ around $b$. Consequently, a small number of training circuits results in similar values of $O^{\rm noisy}$.
Therefore, when we estimate $O^{\rm noisy}$ from a finite shot measurement, a much higher number of shots is required to distinguish them than in the case when they are not clustered around a single value. If the number of shots is not large enough, coefficients of the fitted CDR ansatz are determined randomly by the shot noise making learning the correct CDR ansatz impossible. We find that this toy model explains the poor quality of CDR error mitigation for the case of the 8-qubit XY model and a typical standard training set obtained with small number of shots ($2000$ in this example). As we show in Fig.~\ref{fig:cluster_1obs}(b), error bars of the finite shot estimates of $O^{\rm noisy}$ are so large that one can obtain nearly arbitrary CDR fit coefficients.

\subsection{Distributing the training circuits}
\label{sec:dist}

Despite the observed clustering of $O^{\rm exact}$, near-Clifford training circuits with large $|O^{\rm exact}|$ can usually be obtained. As shown in Appendix~\ref{app:tcirc_sol}, for a circuit of interest compiled to the native IBM's gate set and an observable of interest being an element of Pauli group, a $N=1$ training circuit with a single non-Clifford $R_Z(\theta)$ gate and an arbitrary $ -1 \le O^{\rm exact} \le 1$ can be found, as long as there is at least one $R_Z(\theta)$ such that $O^{\rm exact}$ depends non-trivially on $\theta$. Additionally, if such a training circuit exists, it can be found with a cost that scales linearly with the number of the Clifford gates. Also, Appendix~\ref{app:tcirc_sol} shows that such circuits can be constructed for popular ansaetze as hardware efficient ansatz and Quantum Approximate Optimization Algorithm (QAOA) ansatz, as long as Pauli observable of interest is not an identity.
 
Based on this result, to improve the shot-efficiency of CDR, we propose to use training circuits with a widely spread distribution of $O^{\rm exact}$ that has a significant variance. Here, for a Pauli observable of interest, we use a uniform distribution of $O^{\rm exact}$ with variance in $[0.25, 0.5]$. We find that this approach indeed significantly improves the quality of the CDR correction for modest shot requirements as shown in Fig.~\ref{fig:cluster_1obs}(b) and demonstrated by extensive benchmarks in Section~\ref{sec:ben}.

Several approaches can be used to generate well-distributed training data. First, one can optimize angles of $R_Z(\theta)$ gates under a constraint of $N$ non-Clifford angles to minimize the distance between $O_i^{\rm, exact}$ and its target value. We present this approach in detail in Appendix~\ref{app:tcirc_opt}. For a brick wall arrangement of $L$ layers of CNOTs decorated with random single-qubit gates and $L$ proportional to the qubit number $Q$, in Appendix~\ref{app:tcirc_opt}, we show numerically that this approach requires computational time scaling linearly with the gate number and quadratically with $Q$. 
For this setup, in Appendix~\ref{app:tcirc_opt}, we generate the distributed training circuits for $Q=455$, $L=91$ circuits of interest with tens of thousands of CNOTs and single-qubit gates within a few hours of CPU time. 

Second, one can use Markov Chain Monte Carlo (MCMC) sampling of near-Clifford circuits to generate a desired  $O^{\rm exact}$ distribution. We describe an MCMC algorithm in detail in Appendix~\ref{app:tcirc_MCMC}. Furthermore, one can use a semi-analytic approach of Appendix~\ref{app:tcirc_sol} to efficiently generate $N=1$ training circuits with a cost scaling linearly in the number of Clifford gates. In principle, one can also employ the standard random substitution algorithm supplemented by a postselection. Namely, one could generate near-Clifford circuits randomly substituting non-Clifford $R_Z(\theta)$ in the circuit of interest and accept them as training circuits when they have $O^{\rm exact}$ close enough to the desired one. Because random near-Clifford circuits tend to cluster, a large number of these randomly generated circuits would be needed to find those with the desired distributed observable, which can then be used to generate training data efficiently with fewer shots. Therefore, this approach is inefficient. It is worth noting that none of these approaches requires quantum resources.  

Finally, we note that the original CDR proposal~\cite{czarnik2020error} considered the case of training circuits with a distribution of an observable imposed by MCMC sampling of near-Clifford circuits. Nevertheless, both the motivation for that and the distribution were different than the ones considered here. Specifically, Ref.~\cite{czarnik2020error} considered a special case of a circuit of interest obtained by minimization of a cost function of a Variational Quantum Algorithm. To further bias the training circuits towards such a circuit of interest and maximize the correction quality in the large shot number limit, the approach in Ref.~\cite{czarnik2020error} used near-Clifford circuits that minimize the cost function.

\begin{figure*}[t!]
\begin{center}
\includegraphics[width=0.85\textwidth]{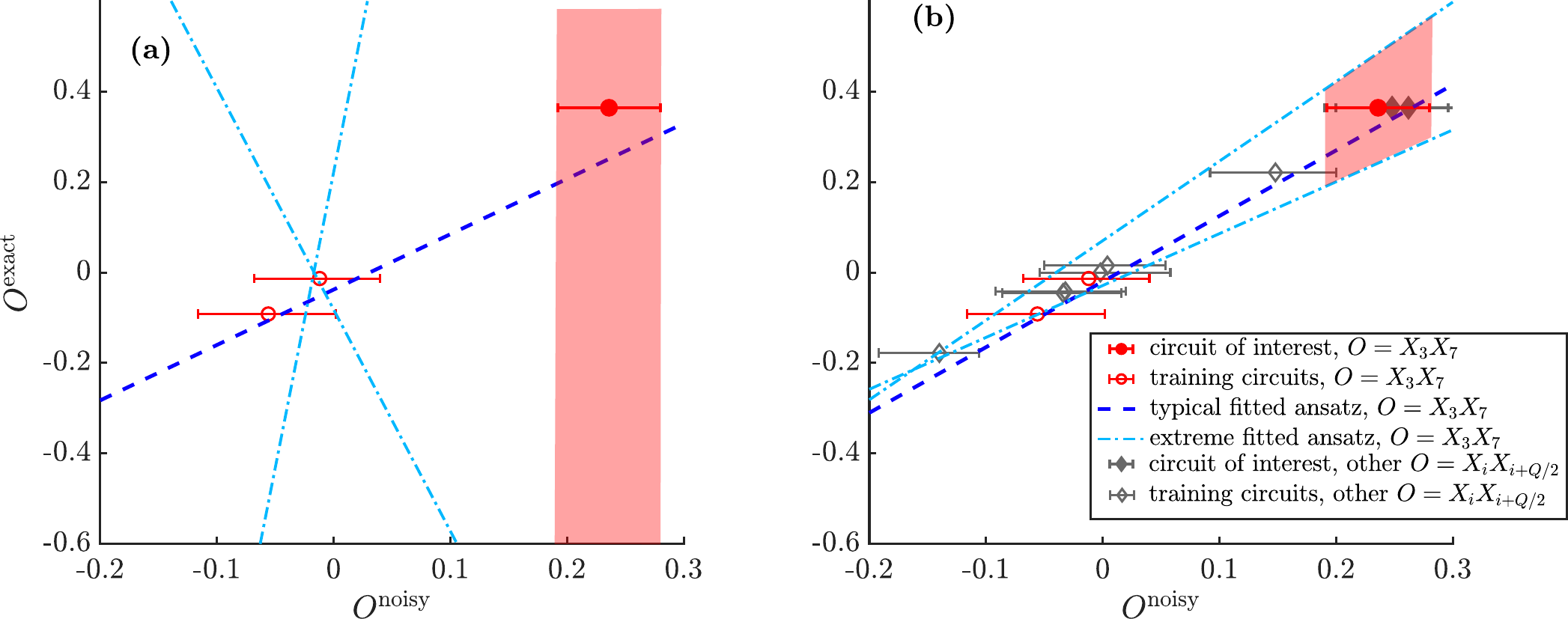}
\end{center}
\caption{ \textbf{Improving CDR shot efficiency by enforcing symmetries of the mitigated circuit.} We perform CDR error mitigation for the ground state of 8-qubit XY model with periodic boundary conditions simulated with the IBM Ourense noise model. We mitigate the expectation value of $O=X_3 X_7$ for the ground state using two training circuits generated by the standard substitution algorithm and performing $1000$ shots for each circuit. In \textbf{(a)}, we show results of standard CDR error mitigation. As in Figure~\ref{fig:cluster_1obs}(b), we find that expectation values of the training circuits (red circles)  cluster around $(0,0)$ and that due to the clustering, the uncertainty of the mitigated expectation value shown by the red shaded area is large. We  visualize this effect performing CDR error mitigation for $100$ instances of shot noise. (The shot noise is represented by the error bars of $O^{\rm noisy}$ in the same way as in Figure~\ref{fig:cluster_1obs}(b)).  We show typical and extreme CDR fits  (defined as in Figure~\ref{fig:cluster_1obs}(b)) with  the blue line and the cyan lines, respectively. In \textbf{(b)}, we take advantage  of the translational invariance of the ground state which implies that  $X_1 X_5$,   $X_2 X_6$, $X_3 X_7$,   $X_4 X_8$ all have the same noiseless expectation values. We measure all of these observables for both our circuit of interest and the training circuits. They are mitigated simultaneously imposing equality of their mitigated expectation values.
The additional observables (induced by symmetry) are shown as gray diamonds in the plot. We correct $100$ different shot noise instances of the noisy expectation values and show the fits corresponding to typical and extremal mitigated expectation value of $X_3 X_7$. We find that   uncertainty of  the mitigated  expectation value  and quality of the typical CDR fit are significantly   improved with respect to \textbf{(a)}. For the sake of transparency we do not show the CDR fits for other observables. They have similar quality to the ones shown here. 
}

\label{fig:cluster}
\end{figure*}

\subsection{Enforcing symmetries of the system}
\label{sec:symmetries}

Another strategy to improve shot efficiency in the presence of clustering is to exploit symmetries of the mitigated circuit. For certain types of symmetric systems, like those with translational symmetry, multiple observables have the same noiseless expectation values. Furthermore, multiple expectation values of such observables can typically be obtained from a single measurement of the circuit. Consequently, in such a case, one can build training data for multiple symmetric observables using the same number of shots as for a single observable. In the presence of such symmetries, we propose to correct all symmetric observables with CDR while enforcing the constraint that all mitigated expectation values are the same. Using expectation values of many symmetric observables, we increase the probability of including data points  with  large $|O^{\rm exact}|$ and   $|O^{\rm noisy}|$ in our training data. The presence of training data with large  $|O^{\rm noisy}|$ improves the shot efficiency of CDR-based error mitigation. By enforcing the constraint, we ensure that such training data affects all mitigated expectation values. 

An example of a circuit to which this approach can  be applied  is  a circuit approximating an eigenstate of a translationally invariant Hamiltonian. For example, in the case of a translationally invariant one-dimensional Hamiltonian with periodic boundary conditions, noiseless observable expectation values 
\begin{displaymath}
X_i X_{i'},  \quad i' = ( i+l) \, {\rm mod} \, Q, \quad  i =1, \dots, Q 
\end{displaymath}
are the same because of symmetry. Here, $Q$ is the number of qubits,  $l$ is an  integer and $X_i$ is a Pauli matrix acting on qubit $i$.  At the same time, for IBM quantum hardware, such expectation values can be measured simultaneously. Therefore, simultaneous CDR error mitigation is possible without requiring additional shot cost. 

 We remark that circuits with symmetries can be obtained through ansaetze respecting symmetries of a problem of interest, including widely-used Quantum Alternating Operator Ansatz~\cite{farhi2014quantum} and Hamiltonian Variational Ansaetze~\cite{wecker2015progress}. Furthermore, systematic procedures to construct symmetric ansaetze for both Variational Quantum Eigensolver and Quantum Machine Learning applications have been proposed~\cite{larocca2022group,meyer2022exploiting,sauvage2022building}. 

More formally, we consider here the case of $M$ observables with equal exact expectation values due to a symmetry of the mitigated system 
\begin{displaymath}
O_1^{\rm exact} = O_2^{\rm exact} = \dots = O_M^{\rm exact}. 
\end{displaymath}
We propose to create training data  by measuring noisy expectation values of all observables for $N_t$ training circuits and classically computing their corresponding exact expectation values 
\begin{displaymath}
\{(O_{ji}^{\rm exact}, O_{ji}^{\rm noisy} ) \}, \,  i = 1, \dots, N_t, \, j = 1, \dots, M. 
\end{displaymath}
In a more general case one can also have  different sets of training circuits for each observable ($N_t$ may depend on observable index $j$).

We propose to find mitigated expectation values of these observables by fitting $M$ linear ansaetze
\begin{equation}
O_j^{\rm exact} = a_j O_j^{\rm noisy} + b_j, \, j = 1,\dots, M
\end{equation} 
to the training data. 
Their  coefficients $a_j,b_j$ are found by minimizing the cost
\begin{equation}
 C = \sum_{ij} (O_{ji}^{\rm exact} - a_j O_{ji}^{\rm noisy} - b_j)^{2}
\label{eq:cost}
\end{equation}
under the constraint that the mitigated expectation values are equal: 
\begin{equation}
a_1 O_{\rm{coi},1}^{\rm noisy} + b_1 = \dots = a_M  O_{\rm{coi},M}^{\rm noisy} + b_M.   
\label{eq:constraint}
\end{equation}
Such a constrained optimization can be performed with the mystic package~\cite{mckerns2011building,mckerns2009mystic}, as in Appendix~\ref{app:symmetry}.   We show an example of the shot-efficiency improvement with this approach in Fig.~\ref{fig:cluster}.

While in this section we concentrate on application of symmetric CDR to improve shot-efficiency of error mitigation, it can be also used to enforce symmetries of the mitigated circuit for the mitigated expectation values. Such symmetries are usually broken by the noise, therefore the standard CDR is not guaranteed to restore them.

\section{Benchmark results }
\label{sec:ben} 
\subsection{The ground state of an XY model }
\label{sec:GS}

\subsubsection{The setup}

We test the proposed methods for the ground state of a one-dimensional XY model with periodic boundary conditions given by the Hamiltonian 
\begin{equation}
H = \sum_{\langle i,j \rangle}  X_i X_j+ Y_i Y_j \ .
\label{eq:H}
\end{equation}
Here $X_i$ and $Y_i$ are Pauli matrices acting on qubit $i$. This Hamiltonian is translationally invariant and it preserves the Hamming weight, i.e. it has a U(1) symmetry.

We use the Variational Quantum Eigensolver~\cite{peruzzo2014variational} to prepare the ground state for $Q=6,8$ lattice sites (or required qubits) with a noiseless classical simulator and enforced translational symmetry. We employ a hardware-efficient ansatz built out of layers of general nearest-neighbor 2-qubit unitaries with a sufficient number of layers (4 for $Q=6$ and 6 for $Q=8$) such that the energy of the prepared state matches the exact ground state energy with accuracy better than $10^{-13}$. The obtained circuits are compiled using the IBM native gate set.   

Here, we reduce the error in the expectation values of long-range correlators for the ground state of the model. Because of the symmetries of the ground state, we have $N$ identical half-chain correlators:  
\begin{equation}
\begin{split}
    O_j &= X_j X_{Q/2+j}, \quad 1 \leq j \leq Q/2, \\
    O_j &= Y_{j-Q/2} Y_{j}, \quad Q/2+1 \leq j \leq Q,
\end{split}    
\label{eq:corr}
\end{equation}
with equal noiseless expectation values.
Below we mitigate the errors of expectation values of these observables distributing the training circuits and using the symmetry properties of the model.  We note that the $Q=6$ ($Q=8$) circuit compiled to IBM's native gate set has $51$ ($144$) CNOTs, with up to $48$ ($132$) CNOTs in a causal cone of a mitigated observable, $20$ ($32$) non-parallelizable layers of CNOTs, and $150$ ($185$) non-Clifford $R_Z(\theta)$ gates. The causal cone of a mitigated observable contains all the gates affecting its expectation value and indicates error mitigation hardness for local noise~\cite{tran2023locality}.

\subsubsection{Implementation on quantum hardware}
\label{sec:Toronto}
We mitigate half-chain correlators (\ref {eq:corr}) of the $Q=6$ XY model (\ref{eq:H}) using IBM's Toronto quantum computer. As a reference, we also perform CDR correction with a standard algorithm described in Section~\ref{sec:CDR} and Appendix~\ref{app:tcirc_rand} using training circuits obtained via random Clifford substitutions. 
We  use a shot-efficient CDR method combining both approaches described in Section~\ref{sec:eff}.  Namely, we perform symmetric CDR for all $O_j$ in (\ref{eq:corr}) following the algorithm described in Section~\ref{sec:symmetries}. Each symmetric observable $O_j$ requires generating independent $N_t^j$ training circuits with widely spread $O_j^{\rm exact}$. These training circuits are generated using the MCMC algorithm described in Appendix~\ref{app:tcirc_MCMC}. We describe the construction of the training sets for both approaches in more detail in Appendix~\ref{app:ts_toronto}. The  shot-frugal approach used here improves CDR shot-efficiency through spreading $O_j^{\rm exact}$, while symmetric CDR restores uniformity of the mitigated correlators. One may consider  generalizations of the applied algorithm which would allow one to generate training circuits with distributions of several $O_j^{\rm exact}$ being widely-spread simultaneously. Such training circuits would improve shot-efficiency of the current approach while increasing classical computational cost of the method. We leave their investigation to future work.   

\begin{figure}[t]
\includegraphics[width=0.9\columnwidth]{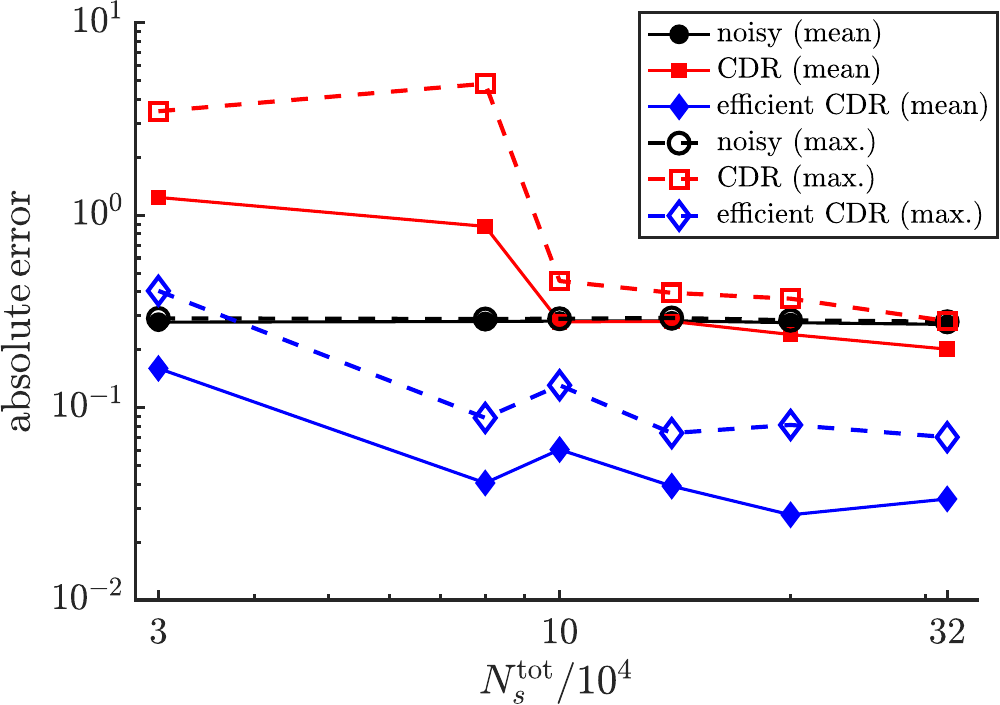}
\caption{ \textbf{Benchmark results for IBM Toronto.} We compare standard and efficient CDR performance for the ground state of the $Q=6$ qubit XY model (\ref{eq:H}) by mitigating symmetric half-chain correlators (\ref{eq:corr}) of the ground state. The comparison is performed for $N_s=10^4$ shots per circuit measurement and $N_t \in \{2,6,8,12,18,30\}$ training circuits. For each $N_t$ and each CDR version we construct 10 independent sets of training circuits, see details in Section~\ref{sec:Toronto} and Appendix~\ref{app:ts_toronto}. Furthermore, for each set we carried out independent measurements of the circuit of interest. For each training set we perform error mitigation and quantify its quality  by computing absolute error of the correlators (\ref{eq:error}). For each method and each $N_t$ value we determine the maximal and the mean absolute error of the mitigated correlators for the sample of training circuits.     We plot these errors versus total shot cost $N_s^{\rm tot}$. For a reference we also plot the errors of  unmitigated, raw results (black lines).    We observe that shot-efficient CDR displays up to an order-of-magnitude correction improvement over the standard method and faster convergence of the mitigated observables with increasing $N_s^{\rm tot}$. 
 }
\label{fig:toronto}
\end{figure}

We compare the performance of the new and standard approaches using IBM's Toronto quantum computer. We note that here, we do not use dedicated measurement error mitigation.
We use $N_s = 10^4$ shots per circuit measurement and compare the performance of both approaches for a total number of training circuits  $N_t \in \{2,6,8,12,18,30\}$. For each value of $N_t$ we generate $10$ different, independent instances of sets of training circuits. The number of non-Clifford gates is set to $N=30$, while the circuit of interest contains 150 non-Clifford gates. We compare the performance by computing the absolute error of the mean of the observable expectation values for each set of training circuits
\begin{equation}
{\rm absolute\, error} = \Big| \sum_j (O_j^{\rm noisy/mitigated} -  O_j^{\rm exact}) \Big| \ . 
\label{eq:error}
\end{equation}
For both methods, we compare mean and maximal absolute errors obtained for a given $N_t$ from a sample of $10$ sets of training circuits. Furthermore, for each set of training circuits we perform independent measurements of the circuit of interest.  We gather the results in Fig.~\ref{fig:toronto} and plot them versus total shot cost of the mitigation, $N^{\rm tot}_s$.

Figure~\ref{fig:toronto} shows data acquired with standard and efficient CDR methods. We concentrate on the realistic regime of $N_s^{\rm tot} \in [3 \cdot 10^4, 3.2 \cdot 10^5]$ and show maximum and mean errors obtained with both methods.   We observe that the new efficient CDR method displays better performance as well as faster convergence with increasing $N_s^{\rm tot}$. These results show a substantial improvement of up to an order of magnitude for the new, efficient method.   Importantly, efficient CDR reaches its correction capabilities already at small shot budgets. This makes it a preferable tool for applications that require the execution of many different quantum circuits. This is a typical situation in many approaches involving variational quantum algorithms.

\begin{figure*}[t]
\centering
\includegraphics[width=0.9\textwidth]{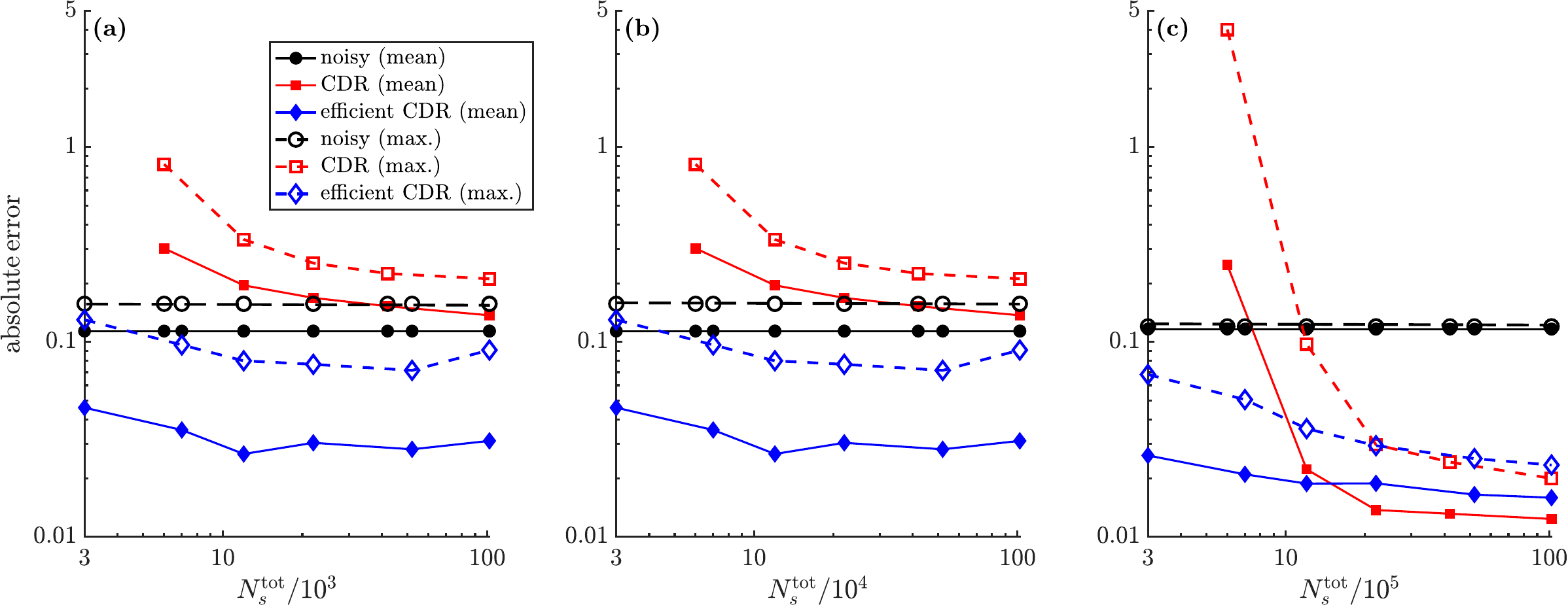}
\caption{ \textbf{Benchmark results for a noise model.} Comparison of the standard (red curves) and efficient (blue curves) CDR for the ground state of $Q=8$ qubit XY model (\ref{eq:H}) by mitigating symmetric half-chain correlators (\ref{eq:corr}) of the ground state. We use a noise model described in detail in Appendix~\ref{app:model}.  The performance is compared for $N_s=10^3$ \textbf{(a)}, $N_s=10^4$ \textbf{(b)} and $N_s=10^5$ shots \textbf{(c)} per a circuit measurement and $N_t \in \{2, \ldots, 100\}$ training circuits. For each $N_t$ and each CDR version we construct $50$ independent sets of training circuits. They are obtained in the same way as for IBM Toronto error mitigation, see Section~\ref{sec:Toronto}. The circuit of interest is measured independently in each set. We plot mean and maximal absolute errors (\ref{eq:error}) of both CDR methods and the noisy data versus total shot cost of the error mitigation $N_s^{\rm tot}$. We observe that the shot-efficient version significantly outperforms the standard one for small and moderate total shot costs $N_s^{\rm tot} = 3\cdot10^3$ to $N_s^{\rm tot} = 1.2\cdot10^6$. 
 }
\label{fig:N8_ns0p1}
\end{figure*}

\subsubsection{Quantum computer noise model}
\label{sec:num}

We perform a more detailed benchmark using a noise model obtained by performing full process tomography of basic gates available on IBM's Ourense quantum computer. Namely, we additionally examine the effect of varying $N_s$ on the performance of shot-efficient CDR. Furthermore, we analyze larger $Q$ and gather better statistics using larger samples of training circuits. The noise model is described in detail in Appendix~\ref{app:model}. Taking into account that the device is an early generation, already retired quantum computer and that we mitigate circuits with larger $Q$ and depth than in the case of our real-hardware implementation, we reduced gate error rates in the model by a factor of $10$ with respect to the original Ourense noise model. This allows us to perform the analysis for noise strengths for which basic CDR provides good quality of error mitigation. While current quantum computers may have noise strengths larger than assumed here, we expect that improvements in quantum computing architectures will reduce the error rates in the near-future.

As in the case of our implementation on IBM's Toronto processor, we mitigate symmetric half-chain correlators (\ref{eq:corr}) of the ground state of the XY model with periodic boundary conditions (\ref{eq:H}). We choose $Q=8$.  We compare the performance of shot-efficient and standard algorithms using the same set-up as for the IBM Toronto benchmark in Section~\ref{sec:Toronto}. This time we consider three different values of $N_s \in \{10^3, 10^4, 10^5\}$ and a larger range of $N_t \in \{2, \ldots, 100\}$. Such choice enables us to test the shot-efficient method for large error mitigation total shot costs. Additionally, for each value of $N_t$ and for each method, we generate $50$ independent sets of training circuits instead of $10$ to decrease finite sample effects. It is currently not possible to perform those extended experiments on real quantum computers due to hardware drift. We discuss those issues in more detail in Appendix~\ref{app:drift}. Effects caused by hardware drift further motivate shot-efficient error mitigation techniques.

We gather the results (the mean and the maximal errors evaluated from a sample of $50$ sets of training circuits) and plot them versus $N_s^{\rm tot}$  in Figure~\ref{fig:N8_ns0p1}. Results obtained with $N_s \in \{10^{3},10^{4}\}$ resemble those observed on the real device. Namely, even in the limit of large $N_t$ we see that the shot-efficient algorithm outperforms standard CDR algorithm. At the same time, we notice that the shot-efficient mitigated results converge much faster with increasing $N_s^{\rm tot}$ than the standard mitigated ones. For $N_s=10^5$ we again see much faster convergence of the shot-efficient CDR which outperforms the standard version for $N_s^{\rm tot} \le 1.2\cdot10^6$. Standard CDR slightly outperforms the shot-efficient version for the largest $N_s^{\rm tot} \ge 2.2\cdot10^6$. 

Overall, we observe that for small and moderate shot costs (up to $1.2\cdot10^6$), the shot-efficient CDR outperforms the standard algorithm. For example, efficient CDR provides improvement of the mitigated results over the noisy ones consistently for all $N_s^{\rm tot} \ge 3\cdot10^3$. This should be compared with standard CDR that shows improvement only for much larger total budgets, $N_s^{\rm tot} \geq 1.2\cdot10^5$. Thus, efficient CDR improves that shot budget threshold by almost two orders of magnitude. Similarly, the shot-efficient method obtains improvement by a factor of at least $4$ with respect to the noisy results for all $N_s^{\rm tot} \ge 7\cdot10^4$. The standard CDR method requires at least $N_s^{\rm tot}=1.2\cdot10^6$ to reach similar improvement.

\begin{figure*}[t]
\centering
\includegraphics[width=0.9\textwidth]{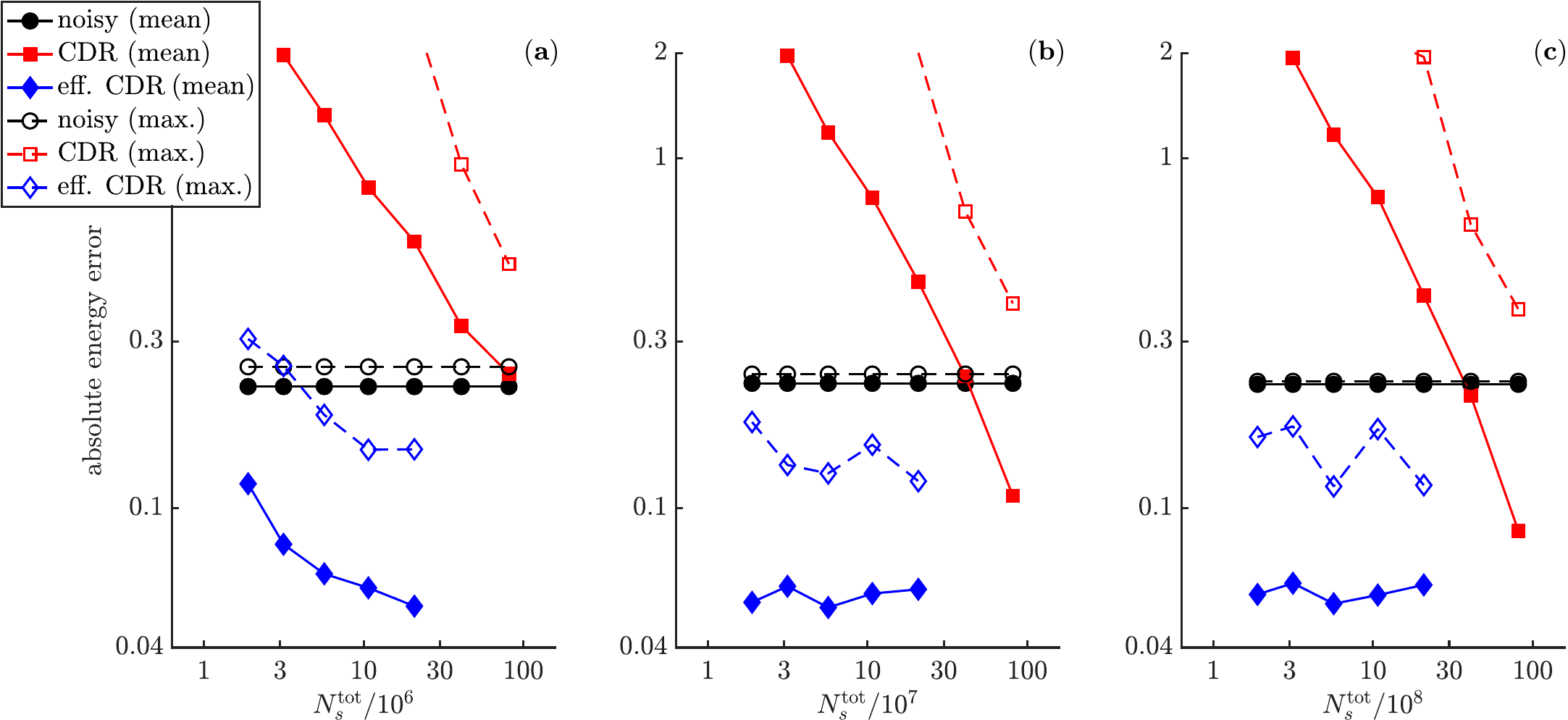}
\caption{ \textbf{Error mitigation of the ground state energy of LiH molecule.} Comparison of the standard and efficient CDR for the energy mitigation of the ground state of LiH molecule. Here, $Q=12$, and we use IBM's Ourense-derived noise model from Appendix~\ref{app:model}. We plot the energy error obtained for $N_t \in \{2, \ldots, 128\}$  ($N_t \in \{2, \ldots, 32\}$) standard (efficient) CDR training circuits per Hamiltonian term. Furthermore, we consider $N_s=10^3$ \textbf{(a)}, $N_s=10^4$ \textbf{(b)} and $N_s=10^5$ shots \textbf{(c)} per $H^{t,{\rm noisy}}_{i}$ evaluation.   
For each $(N_t,\, N_s)$ pair, we perform the standard (the red curves) and shot-efficient CDR (the blue curves)  with $32$  independent training sets and plot the mean (the solid lines)  and maximal (the dashed lines) absolute energy errors~(\ref{eq:en_error}) across the training sets versus the total shot cost $N_s^{\rm tot}$ (\ref{eq:budget_LiH}). As a reference, we show the unmitigated energy errors as the black curves.   
 }
\label{fig:LiH}
\end{figure*}

\subsection{LiH molecule ground state}
\label{sec:LiH}

\subsubsection{The setup}

As a second test case we consider error mitigation of the ground state energy of a LiH (lithium hydride) molecule at the equilibrium bond length 1.55~{\AA}.  An STO-3G basis set is used resulting in a 12-qubit Hamiltonian $H_{\rm LiH}$ which is a linear combination of $n_t=631$ Pauli strings $H^t$
\begin{equation}
H_{\rm LiH} = \sum_{t=1}^{n_t} c_t H^t, 
\end{equation}
with $c_t$ being the coefficients.  We derive the Hamiltonian using \textit{Openfermion} package~\cite{mcclean2017openfermion}. 
We approximate the ground state of the Hamiltonian with a hardware-efficient ansatz built out of  4 layers of general nearest-neighbor 2-qubit unitaries that were found by classical optimization of the ansatz. Subsequently, 2-qubit unitaries were compiled to native IBM gates~\cite{vatan2004optimal} ($R_X(\pi/2)=e^{-i \pi/4 X}$, $R_Z(\theta)=e^{-i \theta/2 Z}$, CNOT) resulting in a circuit with 72 CNOTs, with up to 72 CNOTs in a causal cone of $H^t$, $12$ non-parallelizable layers of CNOTs, and 369 non-Clifford $R_Z(\theta)$ gates. 

\subsubsection{Quantum computer noise model}
\label{sec:LiH_res}

We perform error mitigation using a noise model detailed in Appendix~\ref{app:model} that was derived using IBM's Ourense process tomography. 
We perform CDR for each term of the Hamiltonian $H^t$ separately. As the expectation values of multiple terms of the Hamiltonian typically cannot be measured in the same computational basis, we estimate each  $H^{t,{\rm noisy}}_{\rm coi}$ and  $H^{t,{\rm noisy}}_{i}$ using independent shots. For each expectation value estimation, we use $N_s$ shots.
As the qubit  Hamiltonian $H_{\rm LiH}$ does not have symmetries to exploit, for each $H^t$ we construct efficient CDR training sets with  $H^{t,{\rm exact}}_{i}=-0.5+(i-1)/(N_t-1),\, i \in \{1,\ldots, N_t\}$. We generate a different set of training circuits for each $H_t$. Here, $N_t$ is a number of training circuits per Hamiltonian term. The total shot cost of the energy mitigation is
\begin{equation}
N_s^{\rm tot} = n_t (N_t+1) N_s.
\label{eq:budget_LiH}
\end{equation}
The training circuits are generated by the optimization algorithm described in Appendix~\ref{app:tcirc_opt}, where we use $N=10$. In this case we use the optimization approach because it is more time-efficient than the MCMC approach used for the XY model.

We compare the standard and efficient methods for the task of energy mitigation. We define 
\begin{equation}
{\rm absolute\, error} = \Big| \sum_t c_t (H^{t,\rm noisy/mitigated} -  H^{t,\rm exact})  \Big|.
\label{eq:en_error}
\end{equation}
We evaluate the error for $N_t \in \{2,4,8,16,32\}$ and $N_s \in \{10^3, 10^4, 10^5\}$. For the standard method, we perform mitigation with even larger $N_t \in \{64, 128\}$. We perform error mitigation with $32$ independent sets of the training circuits for each  $N_s$ and $N_t$ pair. Each set of training circuits is obtained by randomly initializing the optimization procedure.  For each $(N_t, N_s)$ pair, we compute the average and the maximal error (\ref{eq:en_error})  across the training circuit sets and plot it versus $N_s^{\rm tot}$ in Figure~\ref{fig:LiH}.

We again find that the efficient method vastly outperforms the standard one. In particular, the efficient method converges quickly with increasing $N_s^{\rm tot}$, unlike the standard one. Moderate $N_s^{\rm tot}=10^7$ is large enough to obtain converged energy error. In contrast, the standard CDR converges slowly and underperforms the efficient one even for the highest considered $N_s^{\rm tot}=10^{10}$. For moderate $N_s^{\rm tot} = 10^6$ to $N_s^{\rm tot} = 10^7$, the efficient mitigation error is an order of magnitude smaller than the standard one. For those $N_s^{\rm tot}$ and $N_s=10^4,10^5$, the efficient approach improves by a factor of 3.7 to 4.3 over the noisy results. Moreover, in Appendix~\ref{app:LiH_Nscal}, we show that the performance of the efficient method does not depend significantly on the choice of $N$.

\section{Conclusions and discussion}
\label{sec:conclusions}
We proposed methods to improve the shot-efficiency of CDR. Our new methods rely on the observation that for large and deep circuits, the expectation values of near-Clifford circuits used by CDR to learn the noise effects cluster around $0$. Because of that effect, a large number of shots is required to distinguish noisy expectation values. Consequently, a large number of shots are necessary to learn the correction successfully.  

We proposed two strategies to alleviate this effect. The first one is to generate a distribution of near-Clifford training circuits with widely distributed exact expectation values for the circuit of interest. We show that such a distribution can be efficiently generated with classical optimization or a semi-analytic procedure without requiring any quantum resources.

The second strategy is to exploit symmetries of the mitigated system by identifying multiple observables which by virtue of the symmetry have the same expectation values and can be measured simultaneously. In the presence of such symmetries, we propose that one performs CDR correction simultaneously for all observables by enforcing the symmetry constraints. This procedure effectively increases the number of available data points to optimize the CDR ansatz. It also restores symmetries of the error mitigated expectation values of the observables.  

We have demonstrated that both strategies can be used to improve CDR shot-efficiency. We combined the strategies and applied them for error mitigation of the ground state of the XY Hamiltonian. This model displays translational and U(1) symmetries. The latter one is responsible for Hamming weight preservation.
Using IBM Toronto we have demonstrated an order-of-magnitude improvement in shot-efficiency relative to standard CDR for moderate total shot costs ($3\cdot10^4$ to $3\cdot10^5$). Furthermore, we analyzed the performance of our shot-efficient method in more detail, using an IBM Ourense-based noise model,  finding consistent improvement of our new method over standard methods for a large range of total error mitigation shot costs ($3\cdot10^3$ to $10^6$). As a second example, we performed LiH ground state energy mitigation. For this task, we demonstrated three orders of magnitude reduction of the shot cost using the IBM Ourense-derived noise model despite the mitigated circuit not having exploitable symmetries.

The clustering problem affects all error mitigation methods that learn from near-Clifford circuits sharing Clifford gate arrangement with a general circuit of interest, including CDR generalizations with multiple noise levels in the training data~\cite{lowe2020unified,bultrini2021unifying}. We expect that our proposed strategies will improve the shot-efficiency of those methods as well. The above techniques have been shown to require more shots than CDR to obtain substantial improvement over CDR. Therefore, improvements of their shot efficiency will be crucial for their practical application. Additionally, an open question is to what extent shot-efficiency of learning-based error mitigation can be improved by combining learning from shot-efficient sets of near-Clifford circuits with machine learning models capable of learning correction for multiple observables and circuits of interest~\cite{liao2023machine}.    Another practically relevant issue that needs further investigation is to what extent the shot-efficient methods' performance can be boosted by combination with specialized measurement error mitigation~\cite{hamilton2020scalable, maciejewski2020mitigation, bravyi2021mitigating} or randomized compiling techniques~\cite{urbanek2021mitigating}.    

Our numerical experiments and analytic arguments demonstrate that for a wide variety of problems, the training circuits can be generated using classical resources that scale polynomially with the number of qubits and the number of gates in the circuit of interest.  Nevertheless, an analysis of the computational time scaling is recommended while applying the methods to new circuit structures.   

Moreover, it will be interesting to study the extent to which the proposed techniques affect the performance of CDR in the limit of large shot number. In particular, the symmetric CDR algorithm, apart from improving the shot-efficiency, restores symmetries of the mitigated system restricting possible error mitigation outcomes. A question arises whether such a restriction can improve the quality of error mitigation for degrees of freedom not fixed by the symmetry. 
Similarly, the usage of distributed training circuits in the limit of large shot number requires further examination. It has been observed that in some cases the use of training circuits with mitigated observable distribution close to the expectation value of the circuit of interest improves the quality of CDR corrections~\cite{czarnik2020error} in comparison to the standard Clifford method. It remains to be seen if this effect occurs for the training circuit distributions proposed here. 

Finally, an interesting question is to what extent an application of the shot-efficient method can improve the quality of training for Variational Quantum Algorithms (VQAs). It has been shown that CDR can improve the trainability of VQAs by reversing distortions of the cost function landscape due to the noise~\cite{wang2021can}. At the same time, it has been shown that for limited shot resources, error mitigation can degrade VQAs performance by increasing the uncertainty of the cost function estimation~\cite{wang2021can}. We expect that by reducing the shot cost of the error mitigation up to orders of magnitude, the proposed method will enable an investigation of those effects with real-world hardware.  

\section{Code availability}

Further implementation details are available from the authors upon request.

\begin{acknowledgments}
We thank Mike Martin for insightful discussions. 
PC acknowledges initial support from Laboratory Directed Research and Development (LDRD) program of Los Alamos National Laboratory (LANL) under project numbers 20190659PRD4 with subsequent support by by the National Science Centre (NCN), Poland under project 2019/35/B/ST3/01028.
MM and 
ATS acknowledge initial support from the Los Alamos National Laboratory (LANL) ASC Beyond Moore's Law project with subsequent support by the Laboratory Directed Research and Development (LDRD) program of Los Alamos National Laboratory under project numbers 20210116DR and 20240712ER.
This material is based upon work supported by the U.S. Department of Energy, Office of Science, National Quantum Information Science Research Centers, Quantum Science Center (LC). LC was also initially supported by the U.S. DOE, Office of Science, Office of Advanced Scientific Computing Research, under the Quantum Computing Application
Teams program.
This research used resources provided by the Los Alamos National Laboratory Institutional Computing Program, which is supported by the U.S. Department of Energy National Nuclear Security Administration under Contract No. 89233218CNA000001.
\end{acknowledgments}
\appendix

\section{Training circuit generation  - standard algorithm. } 
\label{app:tcirc_rand}

Here we outline an algorithm to generate CDR training circuits used as a reference in the main text. The algorithm is based on random substitution  of non-Clifford gates in a circuit of interest by nearby Clifford gates. The algorithm was originally proposed in~\cite{czarnik2020error} and was used and incorporated in subsequent error mitigation frameworks~\cite{lowe2020unified,wang2021can}.
Here we present it for a circuit compiled with native IBM gates $X$, $\sqrt{X}$, $R_Z(\theta)$ and CNOT.

$X$,  $\sqrt{X}$ and CNOT are Clifford gates  while $R_Z(\theta)= e^{-i\theta/2 Z}$ is a  Clifford gate only for $\theta=k\pi/2$, where $k$ is an integer. Therefore, for a circuit containing $\widetilde{N}$ non-Clifford $R_Z$  gates we replace $\widetilde{N}-N$ $R_Z$ gates by Clifford $R_Z$ rotations 
to obtain a training circuit with $N$ non-Clifford gates. We denote angles of the non-Clifford rotations by $\theta_i, i \in \{1,\ldots,\widetilde{N}\}$. 
We replace these gates using the following procedure:
\begin{enumerate}
\item 
Take the circuit of interest as an input circuit. Repeat steps \ref{tr2} and \ref{tr3} $\widetilde{N}-N$ times. 
\item
\label{tr2}
For each non-Clifford gate $R_Z(\theta_i)$  in the input circuit,  calculate weights $w_{ik}=e^{-d_{ik}^2/\sigma^2}$, $k \in \{0,1,2,3\}$. Here, $d_{ik} = ||e^{i \theta_i/2} R_Z(\theta_i) -  e^{i k\pi/4} R_Z(k\pi/2)||_\mathrm{F}$, $||.||_\mathrm{F}$ is the Frobenius matrix  norm and  $\sigma$ is a parameter.  Note that  $d_{ik}$ quantifies the distance between  $R_Z(\theta_i)$ and $R_Z(k\pi/2)$ in a way which does not depend on any global phase introduced by the action of these gates. We choose $\sigma=0.5$ following~\cite{czarnik2020error,lowe2020unified,wang2021can}. 
\item 
\label{tr3}
Replace a gate $R_Z(\theta_i)$ by $R_Z(k\pi/2)$ according to the probability distribution 
\begin{equation}
p_{ik}  = \frac{ w_{ik}}{\sum_{i=1}^{m} \sum_{k=0}^{3} w_{ik}},
\end{equation}
where $m$ is the number of non-Clifford gates in the input circuit, generating an  output circuit. If it contains more than $N$ non-Clifford gates, take it  as an input circuit  for the next repetition of step \ref{tr2}. 
\item
Take the final output circuit as  the  training circuit. 
\end{enumerate}

\section{Existence of well-distributed $N=1$ training circuits }
\label{app:tcirc_sol}

Here, we consider a circuit of interest compiled to CNOTs and general single-qubit unitaries of form 
\begin{equation}
u(\theta_1, \theta_2, \theta_3) = R_Z(\theta_1) \sqrt{X} R_Z(\theta_2) \sqrt{X} R_Z(\theta_3),
\label{eq:u}
\end{equation}
with $\sqrt{X} = R_X(\pi/2) = e^{-i \pi/4 X}$ up to a global phase, and $R_Z(\theta)=e^{-i \theta/2 Z}$.  We note that CNOTs, $R_Z$, and $\sqrt{X}$ are native gates of IBM's quantum computers. 
We consider an observable of interest being a Pauli string $P=\bigotimes_{i=1}^Q P_i$, with, $P_i\in \{ X, Y, Z, I\}$ and $I$ being an identity. In this Appendix, we analyze a distribution of expectation values of near-Clifford CDR training circuits with $N=1$ non-Clifford  $R_Z$ gates for such a circuit of interest. 

We employ here  Heisenberg picture in which a unitary $U$ acts at an observable $O$ as 
\begin{equation}
O \xrightarrow{U} U^{\dag} O U 
\end{equation}
This corresponds to the unitary acting on a state as 
\begin{equation}
|\psi\rangle   \xrightarrow{U} U |\psi\rangle .  
\end{equation}

We first note that a general Pauli string $P$ can be transformed to a Pauli $Z$ string, $P_Z=\bigotimes_{i=1}^Q P^Z_i$, with $P^Z_i\in \{Z, I\}$,  by a shallow unitary built from single-qubit native Clifford gates, as we have,  
\begin{equation}
Y  \xrightarrow{R_Z(\pi) R_X(\pi/2)} Z, 
\end{equation}
\begin{equation}
X  \xrightarrow{R_Z(\pi/2) R_X(\pi/2) } Z.
\end{equation}
Therefore, by applying at the end of the circuit of interest 
$R_Z(\pi/2) \sqrt{X}$ ($R_Z(\pi) \sqrt{X} $) to qubits at which we have $X$ ($Y$) in $P$, we transform $P$ to $P_Z$ with 
$P^Z_i=Z$ if $P_i \in \{X,Y,Z\}$, and $P^Z_i=I$ if $P_i=I$. This basis transformation reduces the case of an arbitrary Pauli string to the special case of Pauli $Z$ string, while adding a small constant overhead in terms of the circuit depth. Furthermore, a total number of CNOTs in the circuit of interest is not changed. 

To establish a general result, we first consider a toy example of $Q=1$, 
\begin{equation}
U = u(\theta_1, \theta_2, \theta_3),
\end{equation}
with $\theta_1, \theta_2, \theta_3 \in [0, 2\pi)$, and $P^Z = Z$.  In the Heisenberg picture, we have 
\begin{equation}
Z \xrightarrow{R_Z(\theta_1)} Z,
\end{equation}
\begin{equation}
Z \xrightarrow{R_X(\pi/2)}  Y
\end{equation}
\begin{equation}
Y \xrightarrow{R_Z(\theta_2)}  \cos(\theta_2) Y + \sin(\theta_2) X,  
\end{equation}
\begin{equation}
\cos(\theta_2) Y + \sin(\theta_2) X \xrightarrow{R_X(\pi/2)}  -\cos(\theta_2) Z + \sin(\theta_2) X,  
\end{equation}
\begin{align}
&-\cos(\theta_2) Z + \sin(\theta_2) X \xrightarrow{R_Z(\theta_3)}  \cos(\theta_2) Z +\\&-
 \sin(\theta_2) (\cos(\theta_3) X  -\sin(\theta_3) Y). \nonumber
\end{align}
Then, 
\begin{equation}
\langle P^Z \rangle = {\rm Tr} \bigg(  |0\rangle\langle 0| \big( -\cos(\theta_2) Z +  \sin(\theta_2) (\cos(\theta_3) X  -\sin(\theta_3) Y)
\big)   \bigg),
\end{equation}
yielding
\begin{equation}
\langle P^Z \rangle = -\cos(\theta_2).
\end{equation}
Therefore, we can set $R_Z(\theta_1)$ and $R_Z(\theta_3)$ to be Cliffords and obtain an arbitrary  $ -1 \le \langle P^Z \rangle \le 1$ by adjusting $\theta_2$.
If $P^Z=I$ instead, $P^Z \xrightarrow{U} P^Z$, and  $\langle P^Z \rangle = 1$.

Next, we note that we can easily generalize this result to the case of the observable of interest  being an arbitrary Pauli $Z$ string $P_Z$
and  $U$  being $u(\theta_1, \theta_2, \theta_3)$  acting at a qubit $j$.
If $P^Z_j=Z$, we obtain
\begin{equation}
 P^Z \xrightarrow{U} -\cos(\theta_2) P^Z + \sin(\theta_2) \cos(\theta_3) \widetilde{P} -
\sin(\theta_2) \sin(\theta_3) \dbtilde{P}, 
\end{equation}
where 
\begin{equation}
\widetilde{P} = X_j \bigotimes_{i \neq j} P^Z_i,
\end{equation}
\begin{equation}
\dbtilde{P} = Y_j \bigotimes_{i \neq j} P^Z_i,
\end{equation}
and 
\begin{equation}
\langle P^Z \rangle = {\rm Tr} ( U^{\dag} P^Z U |\vec{0} \rangle \langle \vec{0}|  )  = -\cos(\theta_2), \
\label{eq:first_u}
\end{equation}
where $| \vec{0} \rangle$ = $\otimes_{i=1}^Q |0\rangle_i$. 
While, if $P^Z_j=I$, then  
\begin{equation}
P^Z \xrightarrow{U} P^Z, 
\end{equation}
\begin{equation}
\langle P^Z \rangle = 1. 
\end{equation}

Furthermore, we note that a CNOT gate maps a $Z$ string into a $Z$ string as 
\begin{equation}
Z_1 \xrightarrow{{\rm CNOT}_{12}} Z_1,
\end{equation}
\begin{equation}
Z_2 \xrightarrow{{\rm CNOT}_{12}} Z_1 Z_2,
\end{equation}
\begin{equation}
Z_1 Z_2 \xrightarrow{{\rm CNOT}_{12}} Z_2.
\end{equation}
Here, ${\rm CNOT}_{12}$ is controlled at qubit 1 and targets  qubit 2. 
We also note that a CNOT acting on the initial state $| \vec{0} \rangle$ does not change it,  and that one can express $I$ (up to a global phase) as $u(\theta_1, \theta_2, \theta_3)$ such that its native gate decomposition (\ref{eq:u}) is built from Clifford gates, e.g., 
\begin{equation}
I \propto u(\pi, \pi, 0). 
\end{equation}

Now, we propose an algorithm to find a near-Clifford training circuit with a single non-Clifford $R_Z$ -- for a circuit of interest compiled to $u$ and CNOT gates and an observable of interest $P^Z\neq I$ -- such that $\langle P^Z \rangle = - \cos(\theta_2)$, $\theta_2 \in [0,2\pi)$.
\begin{enumerate}
\item 
Apply gates in the circuit of interest in the Heisenberg picture until $u$ with a non-trivial dependence of $\langle P^Z \rangle$   on $\theta_2$ (i.e. as in (\ref{eq:first_u})) is found.   
\item 
Construct the training circuit by replacing the found gate by e.g. $u(0,\theta_2,0)$ and all other $u$ gates by  $u(\pi, \pi, 0)$.
\end{enumerate}
Note that for such a choice we can easily evaluate $\langle P^Z \rangle$ by combining Schr\"{o}dinger and Heisenberg picture. Namely we divide the circuit into two parts. The first one is built from the gates preceding $u(0,\theta_2,0)$ in the Heisenberg picture and    $u(0,\theta_2,0)$. It corresponds to an unitary $U_1$. The second one is composed of the rest of the gates and  corresponds to a unitary $U_2$. 
We have 
\begin{equation}
\langle P^Z \rangle =  {\rm Tr} \big(  U_2|0\rangle\langle 0|U_2^{\dag} U_1^{\dag} P^Z U_1 ), 
\end{equation}
but as the gates in $U_2$ do not change the initial state being CNOTs or compilations of identity, we obtain
\begin{equation}
\langle P^Z \rangle = {\rm Tr} \big(  |0\rangle\langle 0| U_1^{\dag} P^Z U_1 \big)= - \cos(\theta_2). 
\end{equation}

We note that in principle for some choices of $P_Z$ and the circuit of interest a $u$ gate with the non-trivial dependence cannot be found and the algorithm fails to produce the training circuit. Nevertheless,  for a non-trival $\langle P^Z \rangle$ that contains $Z$'s  and for popular variational ansaetze the algorithm is guaranteed to produce the training circuit. Examples of such anasetze are  a hardware-efficient ansatz that is  constructed from layers of CNOTs decorated by $u$ gates covering all the qubits and a QAOA ansatz that has  a mixer compiling to a product of $u$ gates acting at all the qubits. The identification of the non-trivial gate can be simplified by simulating a Clifford circuit, which can be performed efficiently using classical computation.~\cite{gottesman1998heisenbergrepresentation}. Finally, we note that usually multiple $u$ gates affecting $\langle P_Z \rangle$ can be found. In such a case, one has a freedom to choose location of the near-Clifford gate or one can attempt to construct $N>1$ training circuits with an arbitrary $\langle P_Z \rangle$, as demonstrated numerically in Appendix~\ref{app:tcirc_eff}.

\section{Shot-efficient training circuits generation}
\label{app:tcirc_eff}

\subsection{Near-Clifford circuits optimization. } 
\label{app:tcirc_opt}

In this section we present an algorithm for fast generation of near-Clifford training circuits. We start with a circuit of interest $\mathcal{C}$ and an observable $O$. The circuit $\mathcal{C}$ is used to prepare state $|\psi_\mathcal{C}\rangle$ on a noisy quantum computer. Error mitigation schemes aim to correct $\langle \psi_\mathcal{C} | O |\psi_\mathcal{C} \rangle$ measured on that computer.

We assume that $\mathcal{C}$ is a brick layer circuit, in which every two-qubit gate has the form $u_j u_{j+1} \mathrm{CNOT}_{j,j+1} $, where
\begin{equation} \label{eq:def_uj}
u_j = R_Z(\theta_{j,1}) \sqrt{X} R_Z(\theta_{j,2}) \sqrt{X} R_Z(\theta_{j,3}) \ .
\end{equation}
Here, $\sqrt{X} = e^{-i \pi/4 X}$ and $R_Z(\theta) = e^{-i\theta/2Z}$. We also assume that $O = Z_k$, where $k$ is a qubit located in the middle of the circuit. The method presented below can be easily extended to more general circuits and other observables.

The CDR method considered in the main text constructs a collection of circuits $\{ \mathcal{C}_i \}$ with the same structure as $\mathcal{C}$ such that $\langle \psi_{\mathcal{C}_i} | O |\psi_{\mathcal{C}_i} \rangle = y_i$. Here, $y_i \in [-1,1]$ are preselected, conveniently distributed values. Circuits $\mathcal{C}_i$ are built in the same way as $\mathcal{C}$. The arrangement of all the gates in $\mathcal{C}_i$ is the same as in $\mathcal{C}$. The circuit differs only by the angles $\vec{\theta}$.

The method consists of two steps: (i) project $\mathcal{C}$ to a near-Clifford circuit $\mathcal{C}_i\vec{\theta})$ and (ii) find $\vec{\theta}$ such that
\begin{equation} \label{eq:cond_an}
|\langle \psi_{\mathcal{C}_i(\vec{\theta})} | O |\psi_{\mathcal{C}_i(\vec{\theta})} \rangle - y_i| < \epsilon
\end{equation}
for a desired precision $\epsilon$.

In step (i), a small number of one-qubit gates $u$ in $\mathcal{C}$ is randomly selected. That set of gates is then expanded by one-qubit gate that acts on qubit $k$ just before the observable $O = Z_k$ is measured on that qubit. Those gates are replaced according to Eq.~\eqref{eq:def_uj} with random angles $\theta \in (0,2\pi)$. All the other one-qubit gates are replaced also according to Eq.~\eqref{eq:def_uj} but with angles $\theta$ chosen from a discrete sets: $\theta_1, \theta_3 \in \{ 0,\pi/2,\pi,3\pi/2 \} $ and $\theta_2 \in \{ 0,\pi \}$. This way, $\mathcal{C}_i(\vec{\theta})$ is a near-Clifford circuit with single, general one-qubit gate placed at the end of the circuit, at a qubit that is later measured. Appendix~\ref{app:tcirc_sol} guarantees that there exists $\vec{\theta}$ such that the condition in Eq.~\eqref{eq:cond_an} is met.

In step (ii), the optimization
\begin{equation} \label{eq:minalpha}
\min_{\vec{\theta}} |\langle \psi_{\mathcal{C}_i(\vec{\theta})} | O |\psi_{\mathcal{C}_i(\vec{\theta})} \rangle - y_i|
\end{equation}
is performed until the value of the optimized function is below predefined $\epsilon$. 

The efficiency of the approach described here relies on the optimization performed in step (ii). The existence of a solution is guaranteed but an estimate of the time that is spent on finding parameters $\vec{\theta}$ in \eqref{eq:minalpha} may be difficult to obtain. We stress that even though parameters $\vec{\theta}$ can be found by following the procedure outlined in Appendix~\ref{app:tcirc_sol}, the error mitigation method needs a diverse set of circuits to work accurately and training circuits should contain more than one non-Clifford gate. In that setting, the numerical optimization over $\vec{\theta}$ becomes necessary. Here, we perform numerical simulations to estimate the time needed to derive a single near-Clifford circuit $\mathcal{C}_n$.

The circuit of interest $\mathcal{C}$ is chosen randomly as described above, with one-qubit gates selected according to Eq.~\eqref{eq:def_uj}. We consider circuits of sizes up to $Q = 455$ qubits. The circuit depth $L$ has been grown linearly with the number of qubits, $L = Q/5$. The largest circuit $\mathcal{C}$ that we analyze here contains over 41,000 CNOT gates and 83,000 one-qubit gates. We are considering an observable $O = Z_{\lfloor Q/2 \rfloor}$ and work with $\epsilon = 10^{-3}$ and $N=10$ non-Clifford gates. We use off-the-shelf optimization methods to optimize over angles $\vec{\theta}$.

\begin{figure}[t!]
\includegraphics[width=\columnwidth]{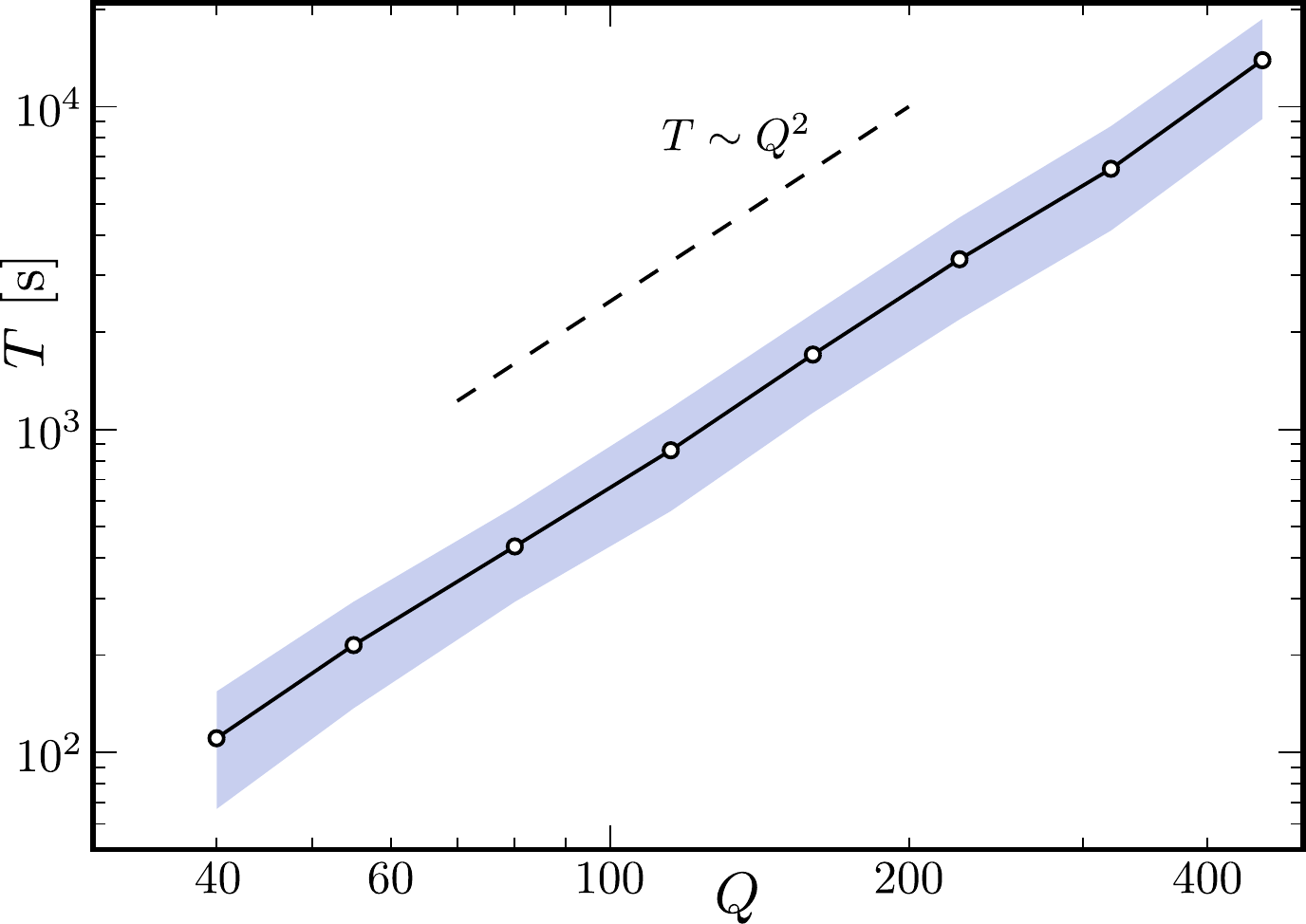}
\caption{ {\bf Computational cost of obtaining the training data.} The plot shows the scaling of the time $T$ to find one near-Clifford $N=10$ circuit $\mathcal{C}_i$ with the number of qubits $Q$. We start from a random circuit whose depth $L$ is linear in $Q$. We find that the computation time $T$ scales as $\sim Q^2$. This scaling closely matches the time to simulate near-Clifford circuit, which indicates that the number of cost function (\ref{eq:minalpha}) evaluations to find optimal $\vec{\theta}$ does not itself scale with $Q$. This leads to an efficient generation of the training data. 
}
\label{fig:tvsq}
\end{figure}

Figure~\ref{fig:tvsq} shows time $T$ to find circuit $\mathcal{C}_i$ as a function of the number of qubits $Q$. The value of $y_i$ in \eqref{eq:minalpha} is chosen randomly from $[-1,1]$ for each $\mathcal{C}_i$. The time is averaged over 200 random circuits $\mathcal{C}$. The optimization is executed on a single core. The solid line shows average time and gray area represents variance in the data. The experimentally obtained time to find solution $T$ scales roughly as $Q^2$, as indicated by dashed line on the plot. This dependence comes from simulation of near-Clifford circuits which scales as the number of gates in the circuit. In our case, that number scales as $Q^2$. Our numerical results thus show that there is no overhead from optimization even in the limit of large system sizes and deep circuits. The simple optimization method used here is sufficient to reliably and quickly find near-Clifford circuits needed for the error mitigation method considered in the main text.

\subsection{Markov Chain Monte Carlo near-Clifford circuit sampling.}
\label{app:tcirc_MCMC}

Here we present an alternative algorithm to numerically generate a distribution of training circuits with widely distributed values of $O^{\rm exact}$ used by the shot-efficient CDR method. The algorithm attempts to generate training circuits with  
values of $O^{\rm exact}$ close to target values $y_1, y_2, \ldots, y_{n}$ using Markov Chain Monte Carlo (MCMC) sampling~\cite{hastings1970monte} of near-Clifford circuits. For each target value, $y_k$, we generate an MCMC chain built from near-Clifford circuits with $N$ non-Clifford gates  using  a  target probability distribution $p \propto  e^{-(O^{\rm exact}-y_i)^2/{\sigma_{\rm MCMC}^2}}$ where $O^{\rm exact}$ is the exact expectation value of the mitigated observable for a near-Clifford circuit and $\sigma_{\rm MCMC}$ is a constant. We select a circuit with $O^{\rm exact}$ close to $y_i$ as a training circuit. 
In this work, $O$ is a Pauli string and we  choose $y_i = -0.5 + \frac{i-1}{n-1}, i \in \{1,2,\ldots,n\}$. For this choice, we have a distribution of target values $|y_n-y_1|=1$ which is comparable to the maximal possible spread of $O^{\rm exact}$ (i.e. $2$). Furthermore, we choose $\sigma_{\rm MCMC}=0.01$. 
 
We use a version of the Metropolis-Hastings algorithm~\cite{hastings1970monte} to generate an MCMC chain for a given $y_i$. We assume here that near-Clifford circuits in the chain can be obtained by replacing most of the non-Clifford gates in the circuit of interest by Clifford gates and consequently the circuit of interest can be obtained by replacing some of the Clifford gates in circuits from the chain by non-Clifford gates. The algorithm is:
\begin{enumerate}
\item
Initialize the chain taking as its  first element a near-Clifford circuit with $N$ non-Clifford gates obtained by substituting  most of non-Clifford gates in the circuit of interest by Clifford gates.     In our implementation, we use an algorithm from Appendix~\ref{app:tcirc_rand} to  generate the first circuit. Repeat steps \ref{modify} and \ref{accept} until a circuit with $O^{\rm exact}$ close enough to $y_i$ is generated. 
\item 
\label{modify}
Prepare a candidate circuit by taking the last circuit of the chain and replacing  $N_g$ randomly chosen non-Clifford gates by Clifford gates and $N_g$ randomly chosen Clifford gates by the corresponding non-Clifford gates of the circuit of interest. Note that the new circuit has $N$ non-Clifford gates which are the same as corresponding non-Clifford gates in the circuit of interest.   
\item
\label{accept}
Compute the ratio $w_{\rm cand}/w_{\rm accept}$, where 
\begin{equation}
w_{\rm cand}=e^{-(y_{\rm{cand}}-y_i)^2/{\sigma_{\rm MCMC}^2}},
\end{equation}
\begin{equation}
w_{\rm accept}=e^{-(y_{\rm{accept}}-y_i)^2/{\sigma_{\rm MCMC}^2}}
\end{equation}
and $y_{\rm cand}$, $y_{\rm accept}$ are $O^{\rm exact}$ for the new circuit and the last circuit of the chain, respectively.  Generate a random number $u \in [0, 1]$. If $u < w_{\rm cand}/w_{\rm accept}$, accept the new circuit as the last circuit of the chain. Otherwise add a copy of the last circuit to the chain. 
\end{enumerate}
In the limit of long MCMC chains this algorithm is guaranteed to generate a distribution of near-Clifford circuits converging to the target distribution $p \propto  e^{-(O^{\rm exact}-y_i)^2/{\sigma_{\rm MCMC}^2}}$~\cite{hastings1970monte} if near-Clifford circuits with $O^{\rm exact}$ close enough to $y_i$ exist. The existence of such circuits for a wide class of ansatzes is guaranteed by Appendix~\ref{app:tcirc_sol}.

We implement step~\ref{modify}  setting $N_g=5$ and replace $R_Z(\gamma)$ by $R_Z(k\pi/2)$ according to a probability distribution 
\begin{equation}
\begin{split}
p_{k} & = \frac{ w_{k}}{ \sum_{k=0}^{3} w_{k}}, \quad w_{k}=e^{-d_{k}^2/\sigma^2}, \\
d_{k}  = || & e^{i \theta/2} R_Z(\theta) -  e^{i k\pi/4} R_Z(k\pi/2)||_\mathrm{F}. 
\end{split}
\end{equation} 
Here, we set $\sigma=0.5$. We find that such implementation results in fast convergence of the MCMC chains to the target distribution.

\section{Implementing symmetry constraints}
\label{app:symmetry}

For a system of $M$ observables,
we implement the constraint (\ref{eq:constraint}) by defining
a \verb|penalty| function:

\begin{verbatim}
def condition(x):
    return std(x[:M] * O_coi + x[M:])**2

@linear_inequality(condition)
def penalty(x):
    return 0.0
\end{verbatim}

where \verb|x[:M]| are the coefficients $a_1,\ldots,a_M$,
\verb|x[M:]| are the coefficients $b_1,\ldots,b_M$,
and \verb|O_coi| is an array of noisy expectation values.

Here, we use \verb|std| from \verb|numpy| for an array-based
calculation of standard deviation, and \verb|linear_inequality|
from \verb|mystic.penalty| to create a penalty function that
scales linearly when \verb|condition| is greater than zero.

To ensure that the \verb|condition| is not violated
during the minimization of the cost function (\ref{eq:cost}),
we convert the \verb|penalty| to a \verb|constraint| operator:
\begin{verbatim}
kwds = dict(nvars=2*M, termination=stop)
constraint = as_constraint(penalty, **kwds)
\end{verbatim}

using \verb|as_constraint| from \verb|mystic.constraints|,
where \verb|stop| is the optimization termination condition
defined by using \verb|ChangeOverGeneration| from
\verb|mystic.termination|:
\begin{verbatim}
args = dict{tolerance=1e-12, generations=60}
stop = ChangeOverGeneration(**args).
\end{verbatim}

The operator \verb|constraint| is then applied during minimization
of the cost function as \verb|x' = constraint(x)|,
to ensure all candidate solutions \verb|x'| for coefficients $a$ and $b$
generated by the optimizer satisfy the constraints. This is done
by passing \verb|constraint| to the \textit{constraints} keyword of the
mystic optimizer used to minimize the cost function.

\section{The noise model}
\label{app:model}
We construct the noise model used in Sections~\ref{sec:num},~\ref{sec:LiH_res} with process matrices obtained by gate set tomography (GST) experiments performed on the IBM Oursense quantum computer. These matrices are given in Appendix B of~\cite{cincio2021machine}, where the details of the experiments can be found. Our goal in Section~\ref{sec:num} is to investigate shot-efficiency gains for deep circuits that are challenging to mitigate with current hardware. To achieve that, we reduce the effective noise rates of the model in comparison to the original process matrices' noise rates. Such a reduction is further justified by the fact that IBM Ourense's noise rates are higher than the error rates of current IBM quantum computers.
Furthermore, in our noise model we assume that the noise rates of gates vary for different qubits and different connections between qubits as in the case of real devices.

More explicitly, in our noise model we assume (following ~\cite{cincio2021machine}) that native gates of the device are $\sqrt{X}=e^{-i \pi/4 X}$, CNOT, identity $I$ and $R_Z(\theta)=e^{-i \theta/2 Z}$. $R_Z$~is assumed to be perfect for IBM quantum computers, as it can be performed "virtually" in a zero-duration fashion without incurring significant error~\cite{mckay2017efficient}.  We construct process matrices of   $I$ and $\sqrt{X}$ as convex combinations of the processes matrices from~\cite{cincio2021machine} $\sqrt{X}^{\rm Ourense}, I^{\rm Ourense}, {\rm CNOT}^{\rm Ourense}$ and the noiseless ones $\sqrt{X}^{\rm perfect}, I^{\rm perfect}, {\rm CNOT}^{\rm perfect}$
\begin{equation}
\sqrt{X}^i = p_i\sqrt{X}^{\rm Ourense} + (1-p_i) \sqrt{X}^{\rm perfect},
\label{eq:RX_PM}
\end{equation}
\begin{equation}
I^i = p_i I^{\rm Ourense} + (1-p_i) I^{\rm perfect},
\label{eq:I_PM}
\end{equation}
\begin{equation}
{\rm CNOT}^{ij}  = p_{ij} {\rm CNOT}^{\rm Ourense} + (1-p_{ij}) {\rm CNOT}^{\rm perfect}.
\label{eq:CNOT_PM}
\end{equation}
Here, indices $i$ and $j$ number qubits and $p_i, p_{ij}$ were drawn randomly from a uniform probability distribution on the interval $[0.05,   0.15]$. 

To quantify quality of a noisy gate $g$ in our model we use average process infidelity defined as 
\begin{align}
&1- F(g^{\rm noisy}, g^{\rm perfect}) = \nonumber \\ &=1 - \int d \psi {\rm Tr} \big[  g^{\rm noisy} (| \psi \rangle \langle \psi |) g^{\rm perfect} (| \psi \rangle \langle \psi |)\big], 
\label{eq:fid}
\end{align}
where  $g^{\rm noisy} (| \psi \rangle \langle \psi |)$ 
 and $g^{\rm perfect} (| \psi \rangle \langle \psi |)$ are states  obtained by action of the noisy  and the noiseless gates at a pure state $| \psi \rangle \langle \psi |$, respectively.  The integral for single-qubit (two-qubit) gates is taken over  single-qubit (two-qubit) Haar measure. The Ourense process matrices  have infidelities~\cite{cincio2021machine}
\begin{equation}
1- F(I^{\rm Ourense}, I^{\rm perfect}) = 2.8 \cdot 10^{-3},
\end{equation} 
\begin{equation}
1- F(\sqrt{X}^{\rm Ourense}, \sqrt{X}^{\rm perfect}) = 8.8 \cdot 10^{-4},
\end{equation}
\begin{equation}
1- F({\rm CNOT}^{\rm Ourense}, {\rm CNOT}^{\rm perfect}) = 1.9 \cdot 10^{-2}. 
\end{equation} 
As the process matrices used (\ref{eq:RX_PM}, \ref{eq:I_PM}, \ref{eq:CNOT_PM}) are convex combinations of the Ourense and noiseless process matrices and fidelity (\ref{eq:fid}) if linear in the left argument, we have:
\begin{equation}
1-F(g^i,g^{\rm perfect}) = p_i(1- F(g^{\rm Ourense}, g^{\rm perfect})),
\end{equation}
\begin{align}
&1-F({\rm CNOT}^{ij},{\rm CNOT}^{\rm perfect}) =\\
& p_{ij}(1-   F({\rm CNOT}^{\rm Ourense}, {\rm CNOT}^{\rm perfect})), \nonumber
\end{align}
with $g$ being $I$ or $\sqrt{X}$. 
Consequently, the model gates infidelities  are on average reduced with respect to IBM's Ourense device by an order of magnitude.

Furthermore, we assume the presence of state preparation error. We assume that the initial state $\rho_0$ is a product state of single-qubit states
\begin{equation}
\rho_0 = \rho_0^1 \otimes \dots \otimes \rho_0^N 
\end{equation}
with 
\begin{equation}
\rho_0^i = p_i \rho_0^{\rm Ourense}+ (1-p_i) \rho_0^{\rm perfect},
\end{equation}
where $\rho_0^{\rm Ourense}$ corresponds to the initial state obtained by GST experiments for IBM Ourense and given in Appendix~B of~\cite{cincio2021machine}, while 
$\rho_0^{\rm perfect}= | 0 \rangle \langle 0 |$
corresponds to perfect state preparation.  In our noise model we neglect measurement error as it can be efficiently mitigated by specialized techniques~\cite{hamilton2020scalable, bravyi2021mitigating,maciejewski2020mitigation}. Probabilities $p_i$ were drawn randomly from a uniform probability distribution on the interval $[0.05, 0.15]$. 

To quantify the noisy state preparation quality we use state infidelity 
\begin{equation}
1 - F(\sigma,\rho) = 1 - \big( {\rm Tr} \sqrt{\sqrt{\rho} \sigma \sqrt{\rho}}\big)^2. 
\label{eq:infid}
\end{equation}
For the Ourense state preparation we have~\cite{cincio2021machine}
\begin{equation}
1 - F(\rho_0^{\rm Ourense}, \rho_0^{\rm perfect}) = 9.7 \cdot 10^{-3},
\end{equation}
while for the noise model we obtain 
\begin{equation}
1 - F(\rho_0^{i}, \rho_0^{\rm perfect}) = 9.7 \cdot 10^{-3} p_i,
\end{equation}
i.e. the model infidelity is on average reduced with respect to the real device  by an order of magnitude.

\section{Training sets from the real-hardware implementation }
\label{app:ts_toronto}

\begin{figure*}[t]
\centering
\includegraphics[width=0.88\textwidth]{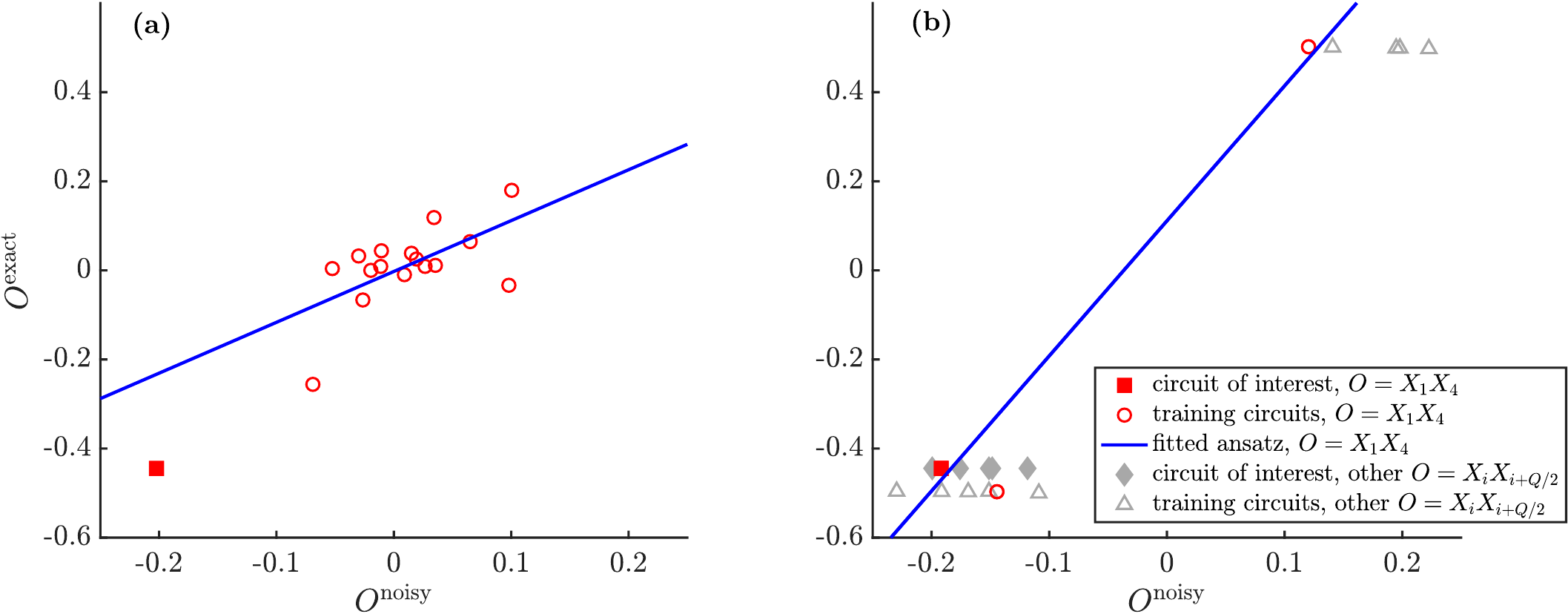}
\caption{ \textbf{Training sets from the real-hardware implementation. } In \textbf{(a)}, we show a standard  CDR training set for the real-hardware implementation from  Figure~\ref{fig:toronto} and the ground state expectation value of $X_1 X_4$  (red square).  Here we use $15$ training circuits (red circles) to correct $X_1X_4$. Number of shots per expectation value evaluation is $N_s=10^4$. Due to clustering of the training circuits quality of the CDR mitigation  is poor as shown by the fitted CDR ansatz (blue line). We obtained similar results for other correlators (\ref{eq:corr}). In \textbf{(b)}, we show a training set for the shot-efficient method and $N_t=12$. Here we correct all symmetric  half-chain correlators (\ref{eq:corr}) using the symmetric CDR algorithm and  generate a widely-spread distribution of the training circuits for each observable independently.  Red squares and circles show  expectation values of $X_1 X_4$ for  the training data and  the circuit of interest. Gray squares and circles show  expectation values of the other symmetric correlators. As in \textbf{(a)}, we have $N_s=10^4$. We obtain good quality of the error mitigation despite total shot cost of the error mitigation  $N_s^{\rm tot}$ being smaller than in \textbf{(a)}. For the sake of transparency we show a CDR fit only for $X_1X_4$. For other correlators (\ref{eq:corr}), quality of the error mitigation is similar.                  }
\label{fig:ts_Toronto}
\end{figure*}

For the standard CDR we  generate training circuits by random substitutions  as described in Appendix~\ref{app:tcirc_rand}. We measure the circuit of interest in one basis to determine expectation values of $O_i=X_i X_{i+Q/2}, i \in \{1,\ldots,Q/2\}$ correlators and in another one to determine expectation values of  $O_{i+Q/2}=Y_i Y_{i+Q/2}, i \in \{1,\ldots,Q/2\}$ correlators. We consider cases of $N_t \in \{2,6,8,12,30\}$ training circuits to correct  $O_i^{\rm noisy}$. To gather statistics for each case we generate $10$ independent sets of training circuits. As we need to measure $X_i X_{i+Q/2}$ and $Y_i Y_{i+Q/2}$ independently,  for $N_t>2$ we measure randomly chosen $N_t/2$ circuits in the $X_i X_{i+Q/2}$ basis and another half of the training circuits in the $Y_i Y_{i+Q/2}$ basis.  This strategy cannot be used for $N_t=2$, so in that case for half of the sets of the training circuits we measure $X_i X_{i+Q/2}$ correlators and for the other half $Y_i Y_{i+Q/2}$ correlators. To randomize the results further, while performing CDR correction for each set of the training circuits, we measure the circuit of interest independently. Figure~\ref{fig:ts_Toronto}(a) shows a typical (resulting in median value of the mitigated error (\ref{eq:error}))  training set generated by that procedure for   $O_1=X_1 X_4$ and $N_t=30$, $N_s=10^4$. Obtained training sets are strongly affected by clustering of the training circuits in the same way as in the case of training sets generated with the noise model in Figure~\ref{fig:cluster_1obs}. That explains poor performance of the standard CDR.

In the case of the shot-efficient version  we generate  distributions of widely-distributed  training  circuits for each $O_i$ independently. To that end we use an algorithm from Appendix~\ref{app:tcirc_MCMC}. In the case of  $N_t \in \{12,18,30\}$ for each $O_i$ we generate  training circuits with values of  $O^{\rm exact}_i\approx-0.5,0.5$, $O^{\rm exact}_i\approx-0.5,0.,0.5$, and $O^{\rm exact}_i\approx-0.5,-0.25, 0, 0.25, 0.5$, respectively.  For $N_t<12$ we choose randomly $N_t/2$ observables $O_i$ for which we generate the training circuits with $O^{\rm exact}_i\approx-0.5,0.5$. To randomize results for each $N_t$ we consider $10$ different sets of training circuits obtained by starting MCMC chains with different initial circuits. Additionally, for each set of training circuits we use independent measurements of the circuit of interest.    We show a typical  training set for $N_t=12$ in Figure~\ref{fig:ts_Toronto}(b).

\section{Effects of hardware drift on performance of learning-based error mitigation}
\label{app:drift}

\begin{figure*}[t]
\centering
\includegraphics[width=0.66\textwidth]{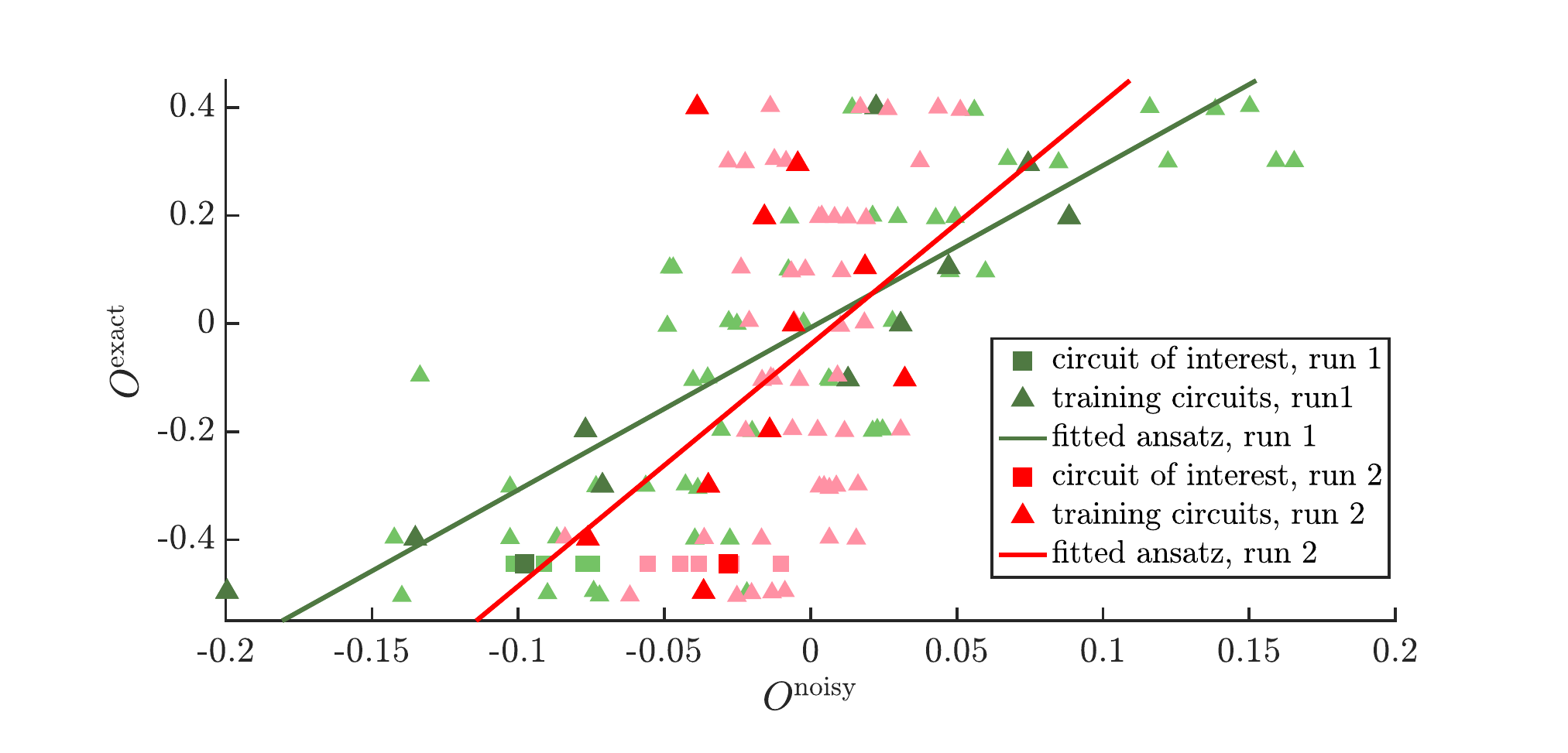}
\caption{ \textbf{Effects of the hardware drift.} We compare two shot-efficient CDR runs conducted for the same circuit of interest at two different times.  We mitigate the half-chain correlators (\ref{eq:corr})  of the  ground state of the 6-qubit XY model using IBM Toronto. Run number 1 was conducted on 01/22/2022 while run number 2 was performed on 03/23/2022. In both runs we used the same training circuits which were compiled to the native gates in the same way. We show here  training sets and fitted ansaetze  from both runs. Green squares and triangles are the  results from the  run number 1 while the red ones are results from run 2. We use dark red and dark green symbols to show  $X_1 X_4$ and  light red and light green ones to plot the  other correlators (\ref{eq:corr}). The drift changed distribution of $O^{\rm noisy}$ affecting parameters of the fitted ansaetze and quality of CDR mitigation.  Here, for the sake of transparency we  plot  only CDR fits (the lines) for $X_1X_4$.  The fits for other correlators give similar  quality of the correction as  the ones shown  here.       For the run 1 we obtain the mitigated error $0.14$ while for the run 2  the mitigated error is $0.28$.  We remark that here quality of the error mitigation is in both cases worse than the one reported in Figures~\ref{fig:toronto},~\ref{fig:ts_Toronto} as  we use  $6$ of $27$ qubits of the device with larger error rates than the ones used in  Figures~\ref{fig:toronto},~\ref{fig:ts_Toronto}.     } 
\label{fig:ts_Toronto_drift}
\end{figure*}

Shot-efficiency of the error mitigation is crucial for obtaining good quality of  quantum computation  when shot resources are limited. Here we argue that shot resources are not only limited by available time of a quantum computer. They are also effectively limited by its variability in time. In the case of the learning-based error mitigation we learn the correction from training circuits assuming that the noise remains constant in time. That assumption is an idealization of real-world devices. In reality, certain characteristics of such devices ``drift'' in time. The shot-efficient method proposed here enables to perform  the error mitigation over a shorter period of time than the standard method. Therefore, it violates the constant noise assumption less severely than the standard CDR method. Consequently, we expect it to deliver better quality of the error mitigation than the standard approach.  While we leave quantifying  this  effect to a future work, we show a proof of principle results demonstrating that the drift affects the quality of the error mitigation. Namely, we performed shot-efficient  CDR error mitigation for the symmetric half-chain correlators~(\ref{eq:corr}) of the ground state of the 6-qubit XY model using IBM Toronto  at two different times. In both cases we  used the same  training circuits which we  compiled to the native gates  in the same way. We found that  coefficients of the fitted CDR ansaetze and average quality of the error mitigation are different for those two runs, see Figure~\ref{fig:ts_Toronto_drift}. For the run number 1 which occurred on 03/23/2022 we obtained the  mitigated error $0.14$. For the run number 2 which was conducted on  01/22/2022  we obtained the  mitigated error $0.28$. Mitigated errors are computed according to~(\ref{eq:error}). That comparison   shows that, indeed,  the device drift can  significantly alter results of the error mitigation. 

\section{Impact of the number of non-Clifford gates in the training circuits, $N$, on the efficiency of CDR performance.   }

\begin{figure*}[tb]
\centering
\includegraphics[width=0.8\textwidth]{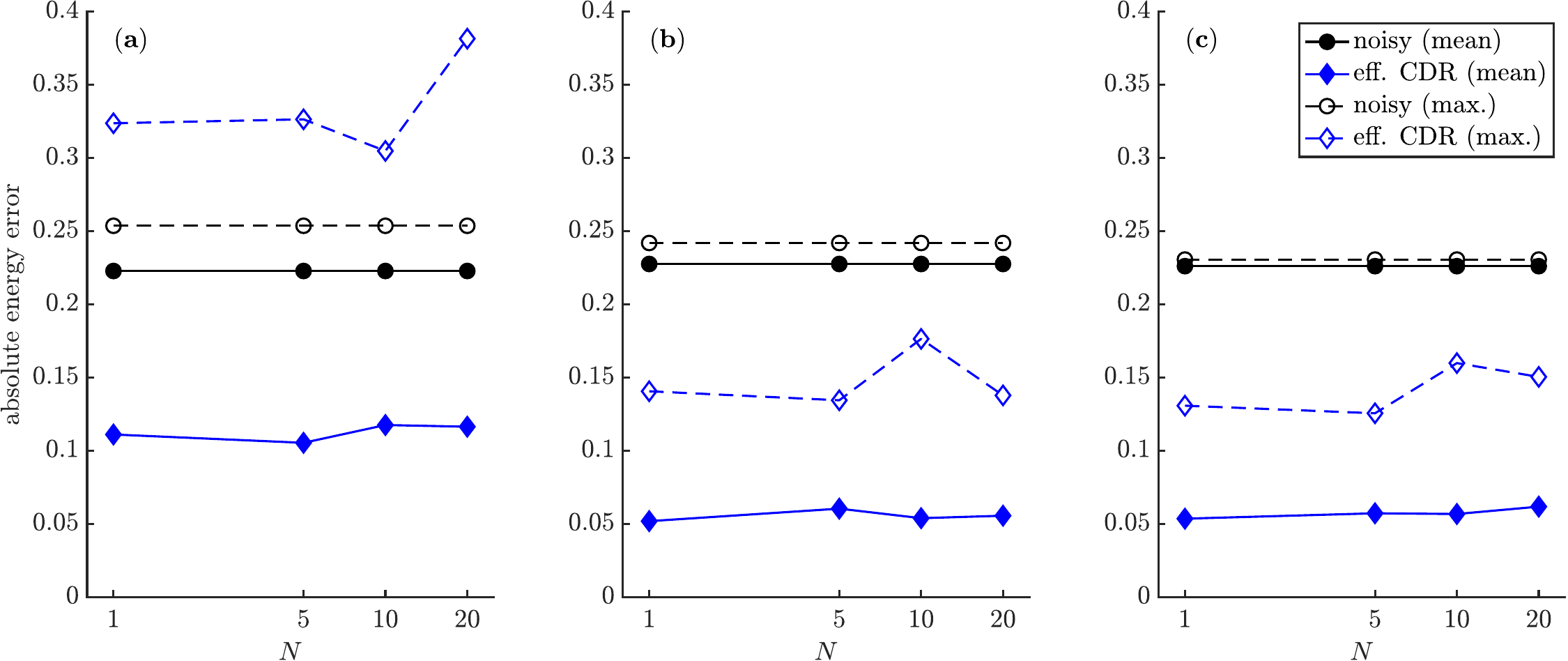}
\caption{ \textbf{Error mitigation of the ground state energy of LiH molecule for different choices of $N$.} The mean and maximal absolute energy errors (\ref{eq:en_error}) of the efficient  CDR method for different choices of a CDR hyperparameter $N\in\{1,5,10,20\}$.  Here, we mitigate the ground state of LiH molecule using IBM's Ourense-derived noise model. The circuit of interest, the noise model, and the training circuit construction method are the same as in Sec.~\ref{sec:LiH}. Here, $N_t=2$, and we show $N_s=10^3, 10^4, 10^5$ results in panels \textbf{(a)}, \textbf{(b}, \textbf{(c)}, respectively. The mean and maximal values are obtained using a sample of  $32$ independent sets of training circuits for each $N$ value. As a reference, we also show errors of the noisy unmitigated data.   
 }
\label{fig:LiH_Nscal}
\end{figure*}

\label{app:LiH_Nscal}

Here, we examine the effects of a choice of a CDR hyperparameter $N$ on the performance of the shot-efficient method for the LiH ground state error mitigation from Section~\ref{sec:LiH}. We choose the same circuit of interest, observable of interest, noise model, and training circuit construction method as in Section~\ref{sec:LiH}.
For such a choice, we compute the absolute mitigated energy error  (\ref{eq:en_error}) for $N\in\{1,5,10,20\}$, as shown in Figure~\ref{fig:LiH_Nscal}. This comparison is performed with $N_t=2$ and $N_s \in \{10^3, 10^4, 10^5$\}. We find that the error remains approximately constant as a function of $N$.

% \clearpage
\bibliographystyle{quantum}

\end{document}